\documentclass[aps]{revtex4}
\usepackage{eurosym}
\usepackage{amsfonts}
\usepackage{amsmath}
\usepackage{amssymb,epsf}
\usepackage{color}
\usepackage{hyperref}
\usepackage{orcidlink}
\usepackage{subcaption}
\usepackage{float}

\begin{document}

\title{Charged Black Holes in Einstein--$U(1)$ Gravity with
Letelier--Alencar Cloud of Strings: Thermodynamics and QPO-Based
Observational Constraints}
\author{B. Eslam Panah\,\orcidlink{0000-0002-1447-3760}}
\email{eslampanah@umz.ac.ir/behzad.eslampanah@gmail.com}
\affiliation{Department of Theoretical Physics, Faculty of Basic Sciences, University of
Mazandaran, P. O. Box 47416-95447, Babolsar, Iran}
\affiliation{Center for Theoretical Physics, Khazar University, 41 Mehseti Str., Baku,
AZ1096, Azerbaijan}
\author{Bidyut Hazarika\, \orcidlink{0009-0007-8817-1945}}
\email{bidyuthazarika1729@gmail.com}
\affiliation{Department of Physics, Dibrugarh University, Dibrugarh, Assam, 786004.}
\author{Prabwal Phukon \orcidlink{0000-0002-4465-7974}$^1$,$^2$}
\email{prabwal@dibru.ac.in}
\affiliation{$^1$ Department of Physics, Dibrugarh University, Dibrugarh, Assam, 786004.\\
$^2$ Theoretical Physics Division, Centre for Atmospheric Studies, Dibrugarh
University, Dibrugarh, Assam, 786004.}

\begin{abstract}
In this study, black hole solutions are derived within the framework of
Einstein gravity, coupled to a $U(1)$ gauge field in the presence of both
the Letelier--Alencar \textit{cloud of strings} and the cosmological
constant. Subsequently, the influence of the \textit{cloud of strings}
parameters and the electric charge on the structure of the event horizon is
investigated. Conserved and thermodynamic quantities associated with these
solutions are computed, and their consistency with the first law of black
hole thermodynamics is verified. To assess the thermodynamic behavior of the
system, the heat capacity and Gibbs potential are derived, thereby enabling
an analysis of local and global stability under variations of the relevant
parameters. Finally, the parameters of the proposed black hole solution are
constrained via a Bayesian Markov Chain Monte Carlo (MCMC) analysis,
utilizing observational quasi-periodic oscillation (QPO) data derived from
stellar-mass, intermediate-mass, and supermassive black holes.
\end{abstract}

\maketitle

\section{Introduction}

Black holes constitute one of the most remarkable and far-reaching
predictions of general relativity, emerging as the final outcome of
gravitational collapse in a broad class of physically relevant scenarios.
Their physical existence has now been firmly established through major
observational breakthroughs. In particular, the detection of gravitational
waves from compact binary mergers by the LIGO--Virgo--KAGRA collaborations
has provided direct evidence for both the formation and coalescence of black
holes, while simultaneously enabling increasingly precise determinations of
their masses and spins \cite{LIGO2016,LIGO2017,LIGO2020}. In parallel,
horizon-scale imaging by the Event Horizon Telescope has opened an
unprecedented observational window onto the strong-gravity environment
surrounding the supermassive compact objects M87$^{*}$ and Sgr A$^{*}$ \cite%
{ETH2019,ETH2022}. Taken together, these advances have substantially
reinforced the motivation for developing and testing black hole solutions
beyond the simplest vacuum geometries.

From an astrophysical perspective, black holes are not expected to exist in
isolation, but rather in environments populated by accreting matter,
magnetic fields, dark matter distributions, or additional long-range fields.
Such external sources modify the background geometry through an effective
stress--energy tensor and may leave observable imprints on the properties of
the compact object. Black holes embedded in nonvacuum environments are
commonly referred to as \textit{dirty black holes} \cite%
{Medved2004,Narayan2005,Cunha2015,Konoplya2021,Cardoso2022,Rahmatov2025,Hoshimov2025,Bakhodirov2025}%
. In this broader setting, the structure of the surrounding matter can play
a central role in determining the horizon configuration, thermodynamic
behavior, dynamical stability, and observational signatures of the spacetime 
\cite{Barausse2014,Macedo2016,Macedo2024}.

Among the matter models considered in this context, the \textit{cloud of
strings} model introduced by Letelier provides a simple and physically
motivated description of a spacetime sourced by a continuous distribution of
one-dimensional strings \cite{Letelier1979}. In its original form, the model
is characterized by an anisotropic stress--energy tensor satisfying 
\begin{equation*}
T^{t}{}_{t}=T^{r}{}_{r}=-\frac{\alpha}{8\pi r^{2}},
\end{equation*}
with vanishing angular components. Because of its analytical tractability
and clear geometric interpretation, this framework has been widely used to
explore the influence of extended anisotropic matter on black hole
spacetimes. Its effects on horizon structure \cite{Rincon2018}, geodesic
motion \cite{Li2021,Belhaj2022,He2022}, and black hole thermodynamics and
phase transitions \cite{Toledo2019,Singh2020,Rodrigues2022b} have been
examined in detail. Related studies have also considered asymptotically flat
and Anti-de Sitter backgrounds \cite{Sadeghi2024}, as well as
generalizations involving higher dimensions \cite%
{Ghosh2014,Toledo2018,Waseem2023}, nonlinear electrodynamics \cite%
{Rodrigues2022,Muniz2025}, and dark energy sectors such as quintessence \cite%
{Mustafa2022,Atamurotov2022}.

A generalized version of the \textit{cloud of strings} model was recently introduced by Alencar \textit{et al.}~\cite{Alencar}. In this framework, the original Letelier configuration was extended through the inclusion of an additional magnetic-like component, $\Sigma_{23}$, which gives rise to a more general anisotropic energy--momentum tensor of the form $\mathrm{diag}(-\rho,-\rho,p,p)$. The resulting matter sector is described by two independent constants and a unique equation of state, while the fundamental geometric properties of the original model, namely spherical symmetry and anisotropy, are preserved. This generalization therefore provides a broader and more flexible setting for exploring the influence of string-like matter distributions on the spacetime geometry, horizon structure, and thermodynamic behavior of black holes. In this context, relativistic tidal effects in an uncharged Letelier--Alencar black hole were examined through geodesic deviation analysis, where significant deviations in the photon sphere, innermost stable circular orbit (ISCO), and tidal-force profiles were identified relative to the Schwarzschild spacetime~\cite{Silva2026}.

An additional motivation for studying such configurations comes from
strong-field observations based on quasi-periodic oscillations (QPOs). QPOs
detected in the X-ray emission of accreting black holes and neutron stars
constitute an important observational probe of the near-horizon region.
These signals appear as nearly periodic modulations of the X-ray flux and
are commonly associated with the dynamics of matter in the inner accretion
disk. Since their first identification in X-ray binaries \cite{17}, a
variety of theoretical models have been developed to explain their origin.
In particular, geodesic models relate the observed frequencies to the
orbital and epicyclic frequencies of test-particle motion in the underlying
spacetime geometry \cite{18,19,20,21,22,23,24,25,26,27,28,29,30,31,32}.
Complementary insight has also been obtained from general relativistic
hydrodynamic and magnetohydrodynamic simulations, which indicate that
accretion-induced plasma instabilities, spiral shock waves, and oscillating
shock cones can generate QPO-like signals \cite{33,34,35,36,37,39}. These
approaches have been successfully applied to several black hole systems,
including GRS~1915+105 \cite{42}, and have also been extended to
supermassive black holes such as M87 \cite{43}. Reviews of particle dynamics
and QPO-related phenomena in black hole spacetimes may be found in Refs.~%
\cite{c1,c2,c3,c4,c5,c6,c7}.

Motivated by these considerations, in this work we investigate charged black
hole solutions in Einstein gravity coupled to a $U(1)$ gauge field, the
generalized Letelier--Alencar cloud of strings, and a cosmological constant.
Our aim is to examine how the combined effects of electric charge, the
cosmological constant, and anisotropic string-like matter modify the
spacetime geometry, horizon structure, conserved quantities, and
thermodynamic behavior of the solution. We further analyze the local and
global stability of the black hole and constrain the model parameters by
means of a Bayesian Markov Chain Monte Carlo analysis using QPO
observations. In this way, the present study provides a unified framework
for connecting the geometric and thermodynamic properties of generalized
dirty black holes with phenomenological constraints from strong-gravity
astrophysical data.

\section{Field Equations and Spacetime}

The Einstein--Maxwell field equations in the presence of a \textit{cloud of
strings}, specifically within the Letelier--Alencar framework, together with
a cosmological constant, can be written as 
\begin{eqnarray}
G_{\mu }^{~\nu }+\Lambda g_{\mu }^{~\nu }-2\left( \frac{g_{\mu }^{~\nu }%
\mathcal{F}}{4}-F_{\mu }^{~\alpha }F_{~\alpha }^{\nu }\right) &=&8\pi T_{\mu
}^{~\nu ^{CS}},  \label{FE} \\
\nabla _{\mu }\left( \sqrt{-g}F^{\mu \nu }\right) &=&0,  \label{Ftr}
\end{eqnarray}%
Here, the Einstein tensor is defined by $G_{\mu}^{~\nu}=\mathcal{R}%
_{\mu}^{~\nu}-\frac{1}{2}g_{\mu}^{~\nu}\mathcal{R}$, where $\mathcal{R}$
denotes the Ricci scalar. In Eq.~(\ref{FE}), $\Lambda$ represents the
cosmological constant, while $g_{\mu}^{~\nu}$ denotes the metric tensor.
Moreover, the Maxwell invariant is given by $\mathcal{F}=F_{\mu\nu}F^{\mu%
\nu} $, with $F_{\mu\nu}=\partial_{\mu}A_{\nu}-\partial_{\nu}A_{\mu}$, being
the Faraday tensor associated with the gauge potential $A_{\mu}$. In Eq.~(%
\ref{Ftr}), the quantity $g$ stands for the determinant of the metric tensor 
$g_{\mu\nu}$, namely $g=\det\left(g_{\mu\nu}\right)$. Throughout this work,
natural units are adopted such that $c=G=1$.

In Eq.~(\ref{FE}), $T_{\mu}^{~\nu^{\mathrm{CS}}}$ denotes the
energy--momentum tensor associated with the Letelier--Alencar \textit{cloud
of strings} configuration \cite{Alencar}. Its nonvanishing components are
expressed as 
\begin{eqnarray}
T_{~~t}^{~t^{CS}} &=&T_{~~r}^{~r^{CS}}=\frac{-g_{s}^{2}\sqrt{l_{s}^{4}+r^{4}}%
}{8\pi r^{4}},  \notag \\
T_{~~\theta }^{~\theta ^{CS}} &=&T_{~~\varphi }^{~\varphi ^{CS}}=\frac{%
g_{s}^{2}l_{s}^{4}}{8\pi r^{4}\sqrt{l_{s}^{4}+r^{4}}},  \label{ComTLA}
\end{eqnarray}%
where $l_{s}$ denotes the characteristic string length, while $g_{s}$
represents the effective coupling parameter that determines the
gravitational strength of the \textit{cloud of strings} sector \cite{Alencar}%
. It is worth emphasizing that, in the limiting case $l_{s}\to 0$ with $%
g_{s}^{2}\to \alpha$, the components of $T_{\mu}^{~\nu^{\mathrm{CS}}}$
consistently recover those of the original Letelier \textit{cloud of strings}
model.

We consider a four-dimensional, static, and spherically symmetric spacetime
described by the following form 
\begin{equation}
ds^{2}=-f\left( r\right) dt^{2}+\frac{dr^{2}}{f\left( r\right) }%
+r^{2}(d\theta ^{2}+\sin ^{2}\theta d\varphi ^{2}),  \label{metric}
\end{equation}
where $f\left( r\right) $ is the metric function.

We now assume the presence of a purely radial electric field, for which the
corresponding gauge potential is taken to be 
\begin{eqnarray}
A_{\mu}=h(r)\,\delta_{\mu}^{t}.
\end{eqnarray}
Substituting this ansatz, together with Eqs.~(\ref{Ftr}) and (\ref{metric}),
into the field equations yields the differential equation 
\begin{eqnarray}
2h^{\prime }\left( r\right) +rh^{\prime \prime }\left( r\right) =0,
\end{eqnarray}
where the prime and double prime denote differentiation with respect to the
radial coordinate $r$. Solving this equation, one obtains 
\begin{eqnarray}
h(r)=-\frac{q}{r},  \label{hr}
\end{eqnarray}
with $q$ being an integration constant associated with the electric charge
of the black hole.

By employing Eqs. (\ref{ComTLA}), (\ref{metric}), and (\ref{hr}) within Eq. (%
\ref{FE}), we can find the components of the equations of motion (Eq. (\ref%
{FE})), which are 
\begin{eqnarray}
eq_{tt} &=&eq_{rr}=\Lambda r^{2}+rf^{\prime }\left( r\right) +f\left(
r\right) -1+\frac{q^{2}}{r^{2}}+\frac{g_{s}^{2}\sqrt{l_{s}^{4}+r^{4}}}{r^{2}}%
,  \label{eqtt} \\
&&  \notag \\
eq_{\theta \theta } &=&eq_{\varphi \varphi }=\left( \frac{f^{\prime \prime
}\left( r\right) }{2}+\Lambda \right) r^{2}+rf^{\prime }\left( r\right) -%
\frac{q^{2}}{r^{2}}-\frac{g_{s}^{2}l_{s}^{4}}{r^{2}\sqrt{l_{s}^{4}+r^{4}}},
\label{eqthethe}
\end{eqnarray}%
where $eq_{tt}$, $eq_{rr}$, $eq_{\theta \theta }$ and $eq_{\varphi \varphi }$
are related to components of $tt$, $rr$, $\theta \theta $ and $\varphi
\varphi $ of the equations of motion (Eq. (\ref{FE})).

\section{Black Hole Solutions}

By considering equations of motion (Eqs. (\ref{eqtt}) and (\ref{eqthethe})),
we can obtain the metric function $f\left( r\right) $ in the following form 
\begin{equation}
f\left( r\right) =1-\frac{2m_{0}}{r}-\frac{\Lambda r^{2}}{3}+\frac{q^{2}}{%
r^{2}}+\frac{g_{s}^{2}l_{s}^{2}}{r^{2}}\mathfrak{F}_{1},  \label{f(r)}
\end{equation}%
where $m_{0}$ is an integration constant related to geometrical mass of the
solution. Also, $\mathfrak{F}_{1}$ is the hypergeometric function in the
following form 
\begin{equation}
\mathfrak{F}_{1}={}_{2}F_{1}\left( \left[ \frac{-1}{2},\frac{-1}{4}\right] ,%
\left[ \frac{3}{4}\right] ,-\frac{r^{4}}{l_{s}^{4}}\right).
\end{equation}

It is notable that the obtained solutions in Eq. (\ref{f(r)}) reduce to the
Reissner-Nordstr\"{o}m AdS black hole solutions when $g_{s}\rightarrow 0$,
i.e., 
\begin{equation}
f\left( r\right) =1-\frac{2m_{0}}{r}-\frac{\Lambda r^{2}}{3}+\frac{q^{2}}{
r^{2}}.  \label{RN}
\end{equation}

To clarify the geometric properties of the spacetime, we first examine the
presence of essential curvature singularities through the evaluation of the
Ricci and Kretschmann scalars. For the class of solutions under
consideration, the Ricci scalar, $\mathcal{R}$, takes the form 
\begin{equation}
\mathcal{R}=4\Lambda +\frac{2g_{s}^{2}\mathfrak{F}_{2}}{l_{s}^{2}}-\frac{%
4g_{s}^{2}r^{4}\mathfrak{F}_{3}}{7l_{s}^{6}},  \label{R}
\end{equation}%
where $\mathfrak{F}_{2}$ and $\mathfrak{F}_{3}$ are introduced as 
\begin{eqnarray}
\mathfrak{F}_{2} &=&{}_{2}F_{1}\left( \left[ \frac{1}{2},\frac{3}{4}\right] ,%
\left[ \frac{7}{4}\right] ,-\frac{r^{4}}{l_{s}^{4}}\right) ,  \notag \\
&&  \notag \\
\mathfrak{F}_{3} &=&{}_{2}F_{1}\left( \left[ \frac{3}{2},\frac{7}{4}\right] ,%
\left[ \frac{11}{4}\right] ,-\frac{r^{4}}{l_{s}^{4}}\right) .
\end{eqnarray}

The Kretschmann scalar ($K=\mathcal{R}_{\mu \nu \rho \sigma }\mathcal{R}%
^{\mu \nu \rho \sigma }$) of this spacetime is%
\begin{eqnarray}
K &=&\frac{16g_{s}^{4}r^{8}\mathfrak{F}_{3}^{2}}{49l_{s}^{12}}+\frac{%
16g_{s}^{2}r^{4}\left( g_{s}^{2}\mathfrak{F}_{2}-\Lambda l_{s}^{2}\right) 
\mathfrak{F}_{3}}{21l_{s}^{8}}-\frac{32m_{0}g_{s}^{2}r\mathfrak{F}_{3}}{%
7l_{s}^{6}}  \notag \\
&&  \notag \\
&&+\frac{4}{63l_{s}^{6}}\left( 108q^{2}g_{s}^{2}\mathfrak{F}_{3}+42\Lambda
g_{s}^{2}l_{s}^{4}\mathfrak{F}_{2}+108g_{s}^{4}l_{s}^{2}\mathfrak{F}_{1}%
\mathfrak{F}_{3}+42\Lambda ^{2}l_{s}^{6}+35g_{s}^{4}l_{s}^{2}\mathfrak{F}%
_{2}^{2}\right)  \notag \\
&&  \notag \\
&&-\frac{16m_{0}g_{s}^{2}\mathfrak{F}_{2}}{l_{s}^{2}r^{3}}+\frac{56g_{s}^{2}%
\mathfrak{F}_{2}}{3l_{s}^{2}r^{4}}\left( q^{2}+g_{s}^{2}l_{s}^{2}\mathfrak{F}%
_{1}\right) +\frac{48m_{0}^{2}}{r^{6}}-\frac{96m_{0}\left(
q^{2}+g_{s}^{2}l_{s}^{2}\mathfrak{F}_{1}\right) }{r^{7}}+\frac{56\left(
q^{2}+g_{s}^{2}l_{s}^{2}\mathfrak{F}_{1}\right) ^{2}}{r^{8}},  \label{K}
\end{eqnarray}%
where the Kretschmann scalar diverges at $r=0$, i.e., $\underset{%
r\rightarrow 0}{\lim }R_{\mu \nu \rho \sigma }R^{\mu \nu \rho \sigma
}\rightarrow \infty $. \bigskip So there is an essential curvature
singularity at $r=0$.

On the other hand, the asymptotical behavior of these solutions is not
(A)dS, because $\mathcal{R}$ and $\mathcal{R}_{\mu \nu \rho \sigma }\mathcal{%
R}^{\mu \nu \rho \sigma }$ are not $4\Lambda $, and $\frac{8\Lambda ^{2}}{3}$
when $r\rightarrow \infty $. This is due to the existence of
Letelier-Alencar \textit{cloud of strings.}

\begin{figure}[tbph]
\centering
\includegraphics[width=55mm]{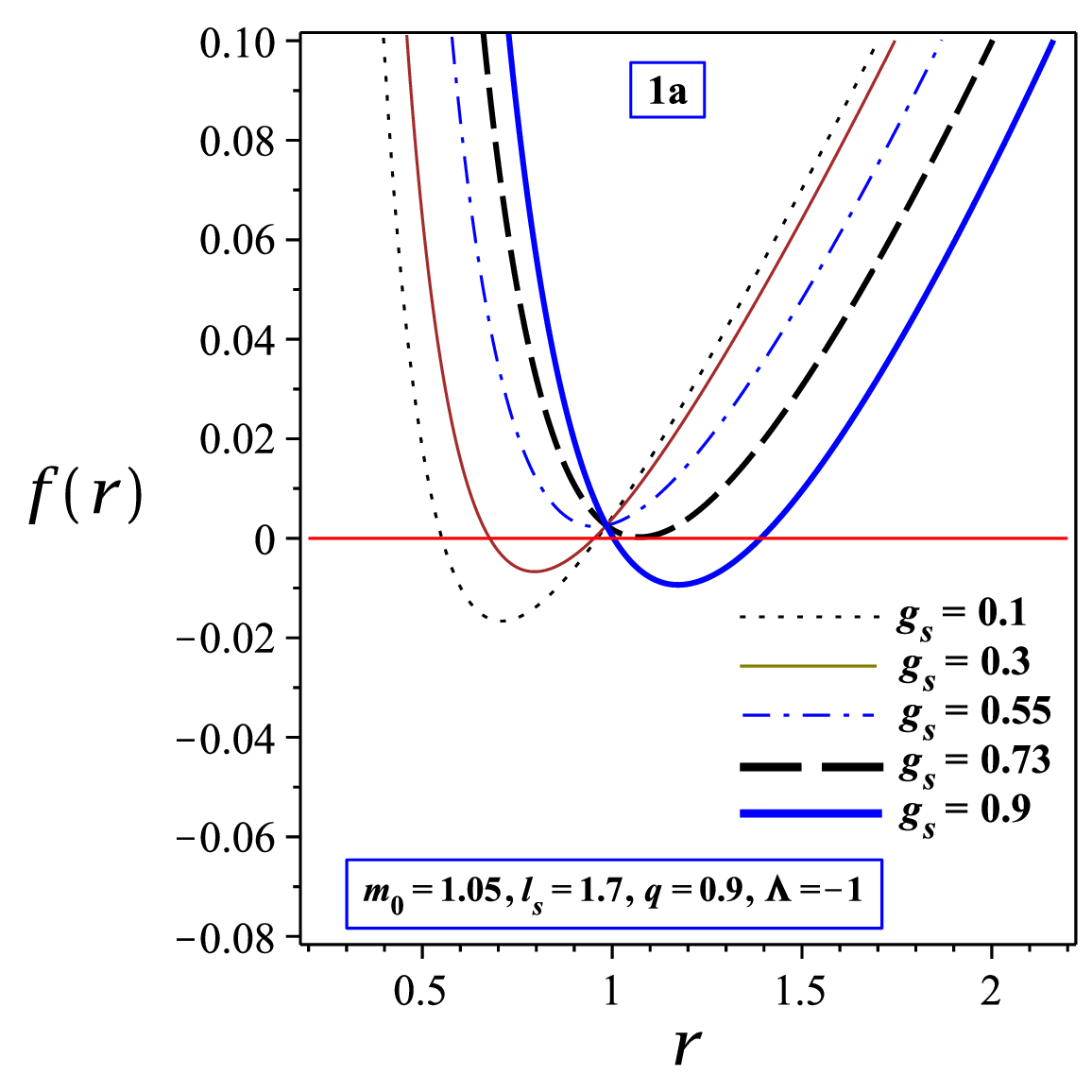} \includegraphics[width=55mm]{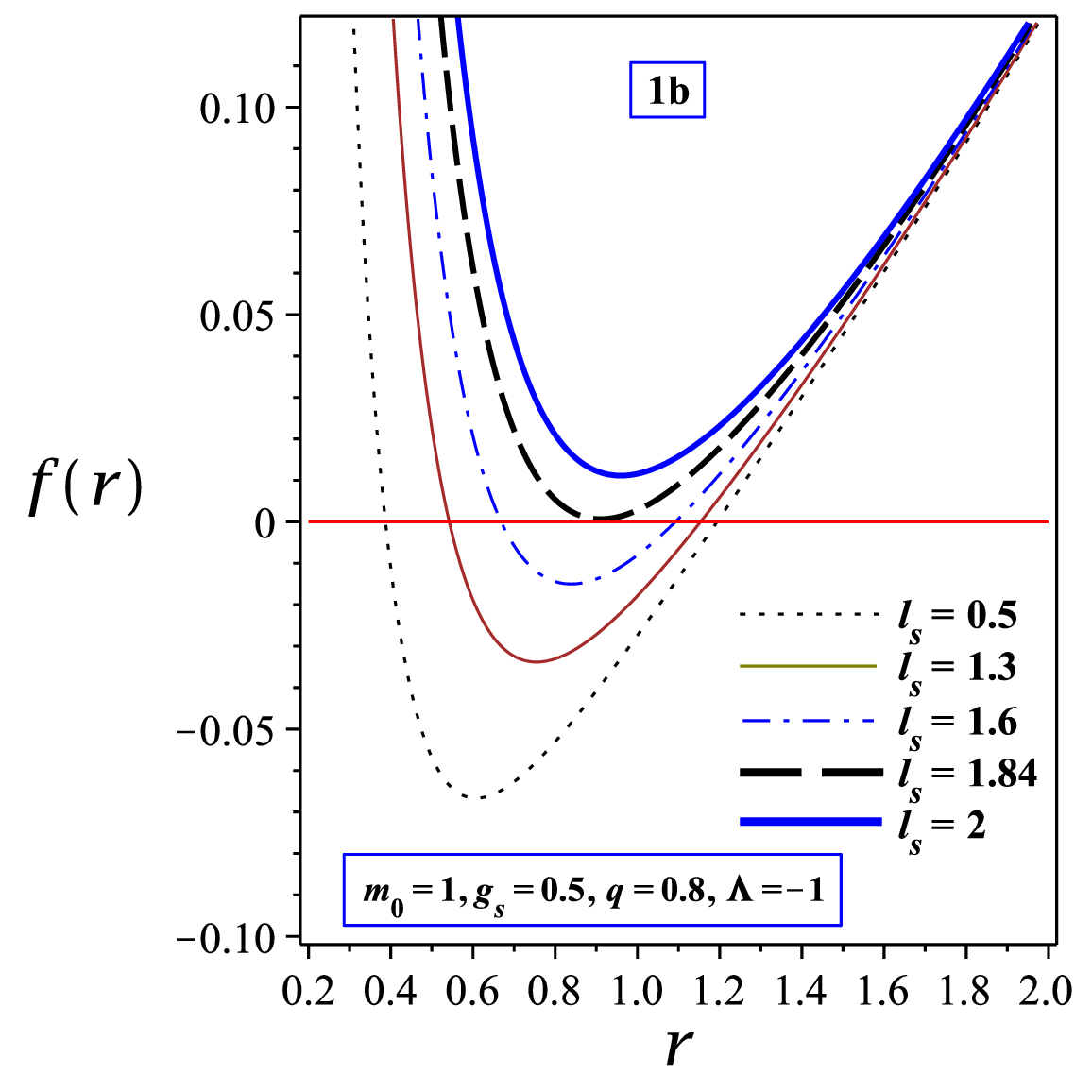} %
\includegraphics[width=55mm]{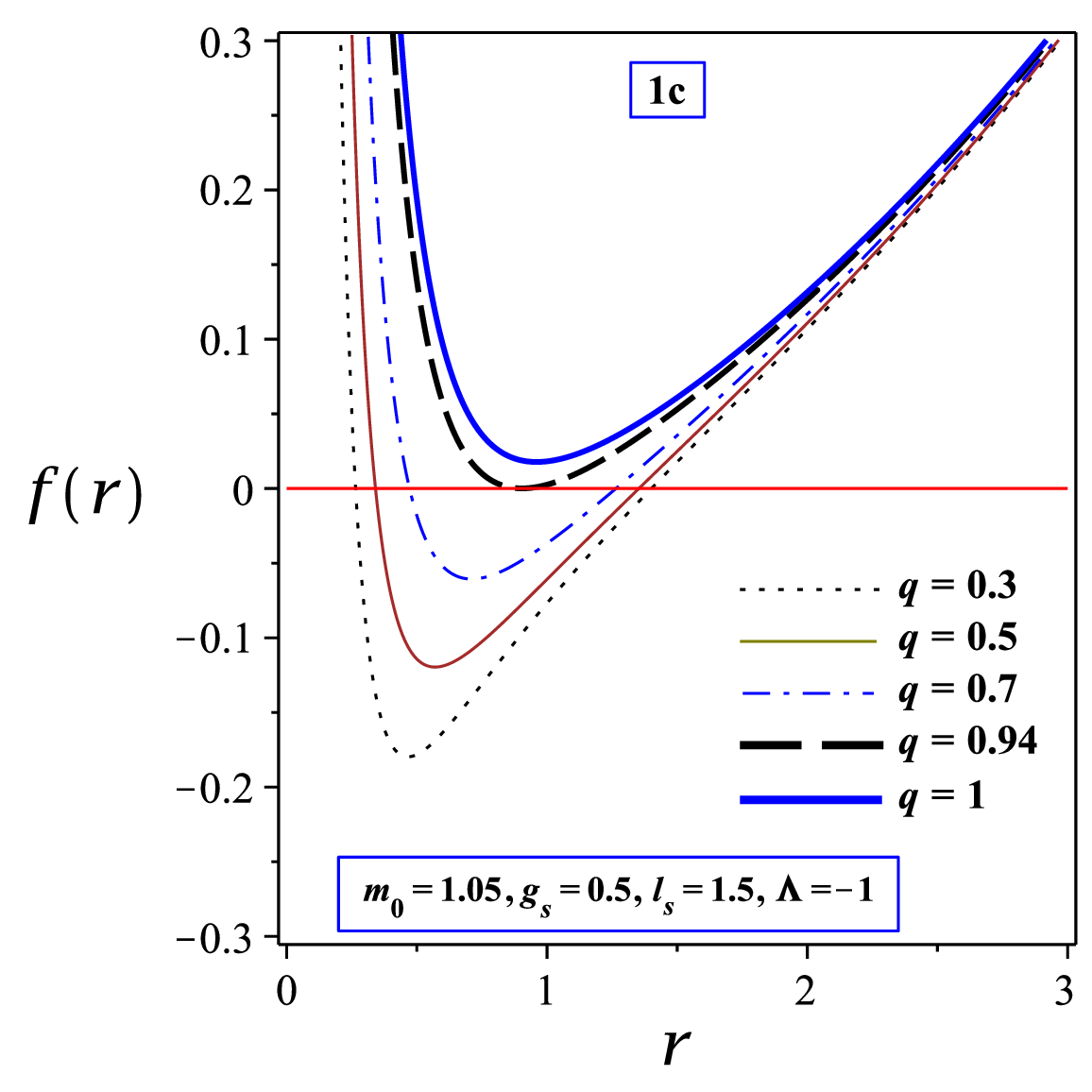}
\caption{The mertic function $f(r)$ versus $r$ for different values of
parameters $g_{s}$ (\protect\ref{fig1}a), $l_{s}$ (\protect\ref{fig1}b), and 
$q$ (\protect\ref{fig1}c).}
\label{fig1}
\end{figure}

The analysis demonstrates that an essential curvature singularity is located
at $r=0$. This singularity is enclosed by an event horizon, as verified by
the behavior of the metric function $f(r)$ shown in Fig.~\ref{fig1}. The
influence of the Letelier--Alencar \textit{cloud of strings} parameters and
the electric charge on the roots of the metric function is then examined. It
is found that three critical values, associated with the parameters $g_{s}$, 
$l_{s}$, and $q$, can be identified and denoted by $g_{s_{\mathrm{crit}}}$, $%
l_{s_{\mathrm{crit}}}$, and $q_{\mathrm{crit}}$, respectively. By varying
these parameters, the number and nature of the roots of the metric function
are modified. The main results can be summarized as follows:

\begin{itemize}
\item[\textbf{i)}] \textbf{ Fig. \ref{fig1}a:} For sufficiently small values of $g_{s}$, two roots are obtained (see the black dotted line). The smaller root is
associated with the Cauchy horizon, whereas the larger root corresponds to
the event horizon. As $g_{s}$ is increased, the radius of the event horizon
is reduced. At $g_{s}=g_{s_{\mathrm{crit}}}$, no root is obtained and a
naked singularity is produced (see the blue dash-dotted line). When $g_{s}$
is increased beyond $g_{s_{\mathrm{crit}}}$, an extremal horizon first
appears (see the black dashed line), and for sufficiently large values of $%
g_{s}$, two roots re-emerge, corresponding again to the Cauchy and event
horizons (see the blue thick line). It should be emphasized that this
behavior differs from that reported in Refs. \cite{Alencar,Silva2026}, which
may be attributed to the presence of the electric charge.

\item[\textbf{ii)}] \textbf{Fig. \ref{fig1}b:} For $l_{s}<l_{s_{\mathrm{crit}}}$, both an
event horizon and a Cauchy horizon are present, and the size of the event
horizon is reduced as $l_{s}$ is increased. For $l_{s}=l_{s_{\mathrm{crit}}}$%
, an extremal horizon is obtained for the solution in Eq. (\ref{f(r)}) (see
the black dashed line). For $l_{s}>l_{s_{\mathrm{crit}}}$, no horizon is
formed and a naked singularity is obtained (see the blue thick line).

\item[\textbf{iii)}] \textbf{ Fig. \ref{fig1}c:} For sufficiently small values of $q$, or
equivalently for $q<q_{\mathrm{crit}}$, two roots are exhibited by the
metric function, corresponding to the event horizon and the Cauchy horizon
(see the black dotted line). Moreover, the event horizon radius is reduced
as the electric charge is increased. At $q=q_{\mathrm{crit}}$, a single root
is obtained, corresponding to an extremal horizon (see the black dashed
line). For $q>q_{\mathrm{crit}}$, no root exists (see the blue thick line).
This qualitative behavior is consistent with that known for Reissner--Nordstr%
\"{o}m black holes.
\end{itemize}

In summary, two roots can be admitted by the metric function when the
electric charge $q$ and the parameter $l_{s}$ take relatively small values,
provided that $g_{s}$ lies in either a sufficiently small or sufficiently
large regime. Furthermore, larger event horizons are associated with black
hole configurations characterized by large $g_{s}$ and comparatively small
values of $q$ and $l_{s}$.

\section{Thermodynamic}

Here, we calculate the thermodynamic quantities for charged Letelier$-$%
Alencar black holes in the presence of the cosmological constant, examining
the influence of the string cloud and charge parameters on these quantities.
The results confirm adherence to the first law of thermodynamics.

Due to the fact that employed metric only contains one temporal killing
vector ($\chi _{\mu }=\left( 1,0,0,0\right) $), one can use the concept of
surface gravity ($\kappa =\sqrt{\nabla _{\mu }\chi _{\nu }\nabla ^{\mu }\chi
^{\nu }}$) for calculating the temperature on the event horizon ($r_{+}$)
which leads to 
\begin{equation}
T=\frac{\kappa }{2\pi }=\frac{1}{2\pi }\sqrt{\nabla _{\mu }\chi _{\nu
}\nabla ^{\mu }\chi ^{\nu }}=\frac{1}{4\pi }\frac{df\left( r\right) }{dr}%
|_{r=r_{+}},  \label{Temp1}
\end{equation}%
by considering Eqs. (\ref{f(r)}) and (\ref{Temp1}), we can get the Hawking
temperature for these black holes as 
\begin{equation}
T=\frac{1}{4\pi r_{+}}-\frac{q^{2}}{4\pi r_{+}^{3}}-\frac{\Lambda r_{+}}{%
4\pi }-\frac{g_{s}^{2}}{12\pi r_{+}^{3}l_{s}^{2}}\left( 3l_{s}^{4}\mathfrak{F%
}_{1_{+}}+2r_{+}^{4}\mathfrak{F}_{2_{+}}\right) ,  \label{Temp2}
\end{equation}%
where $\mathfrak{F}_{1_{+}}=\left. \mathfrak{F}_{1}\right\vert _{r=r_{+}}$
and $\mathfrak{F}_{2_{+}}=\left. \mathfrak{F}_{2}\right\vert _{r=r_{+}}$.

\begin{figure}[h]
\centering
\includegraphics[width=55mm]{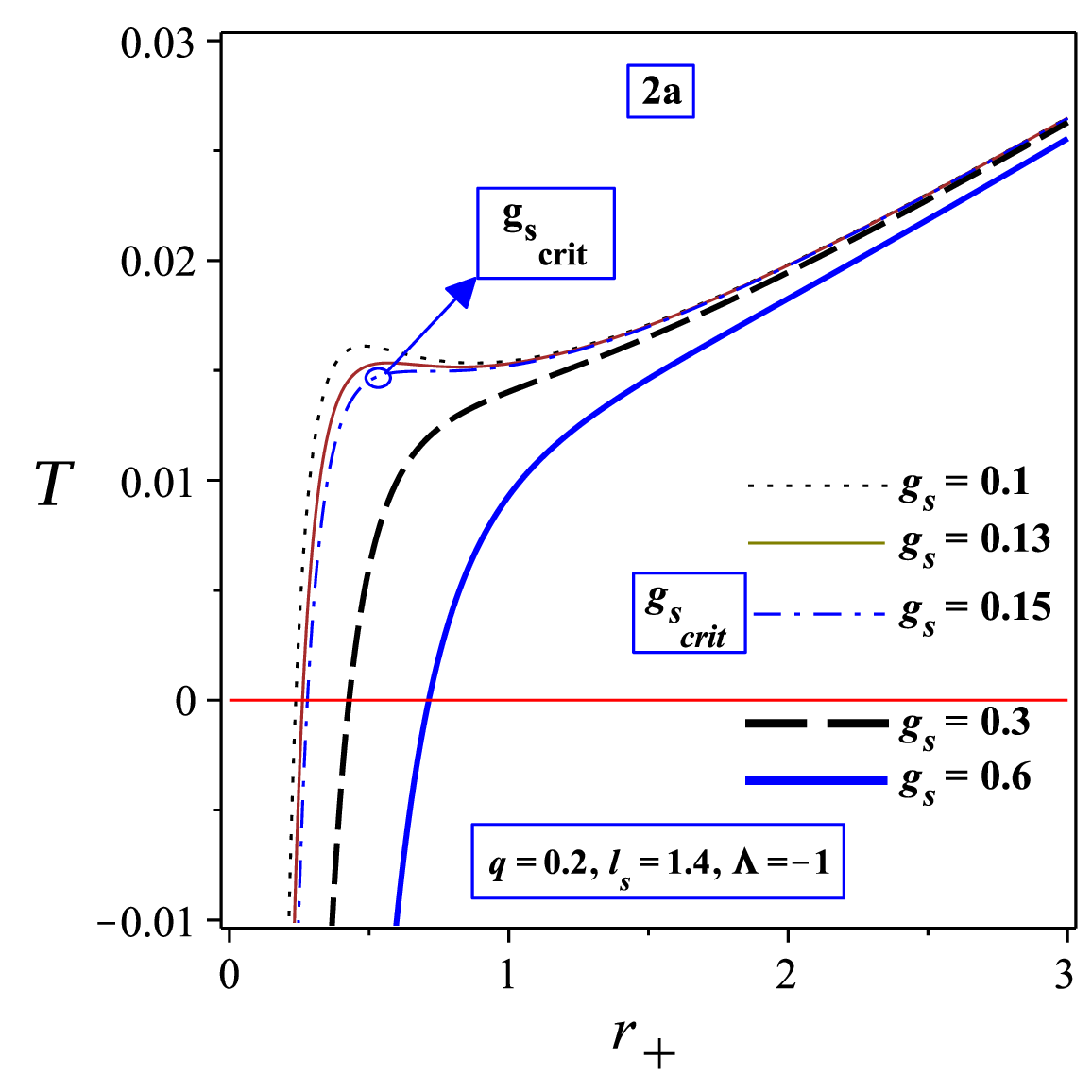} \includegraphics[width=55mm]{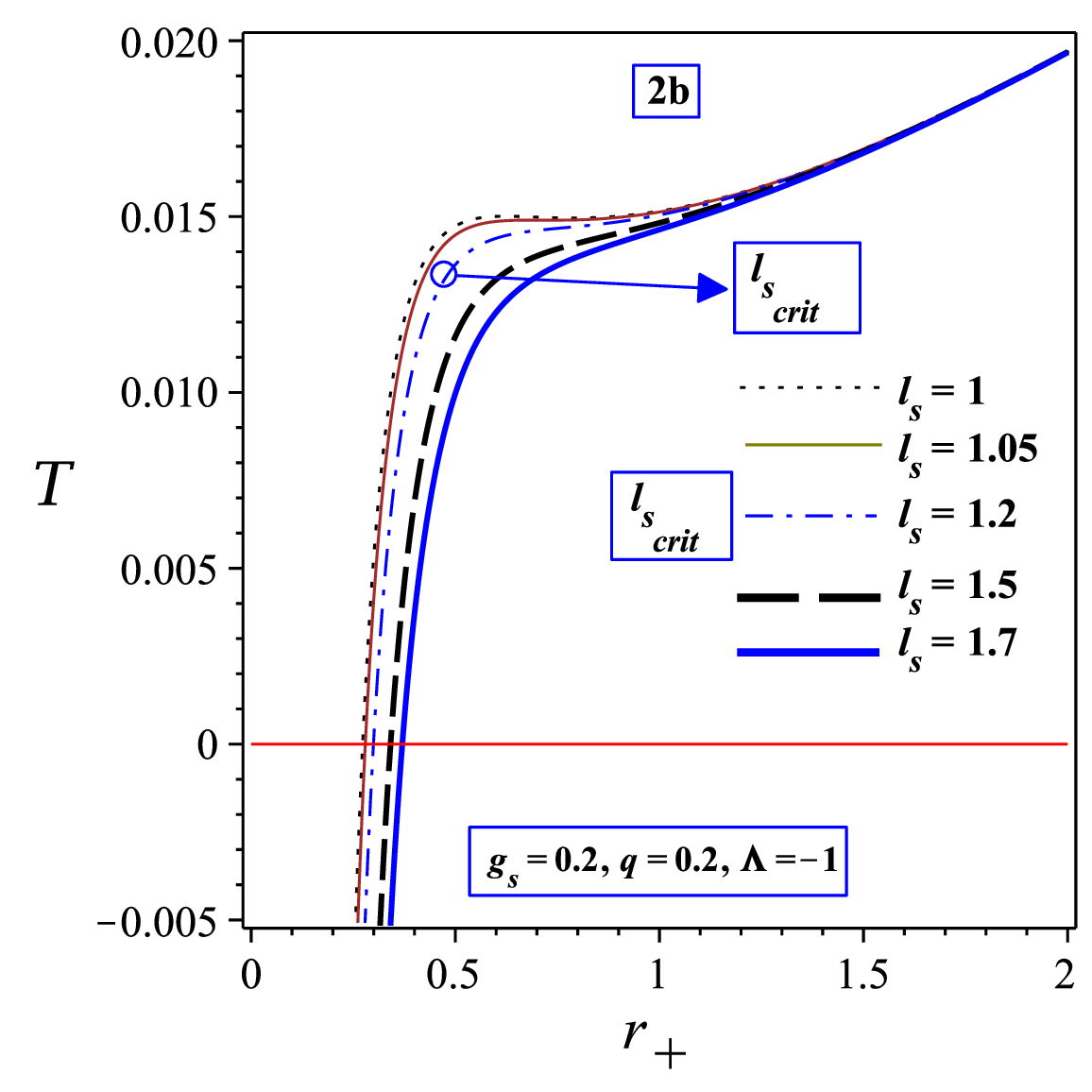} %
\includegraphics[width=55mm]{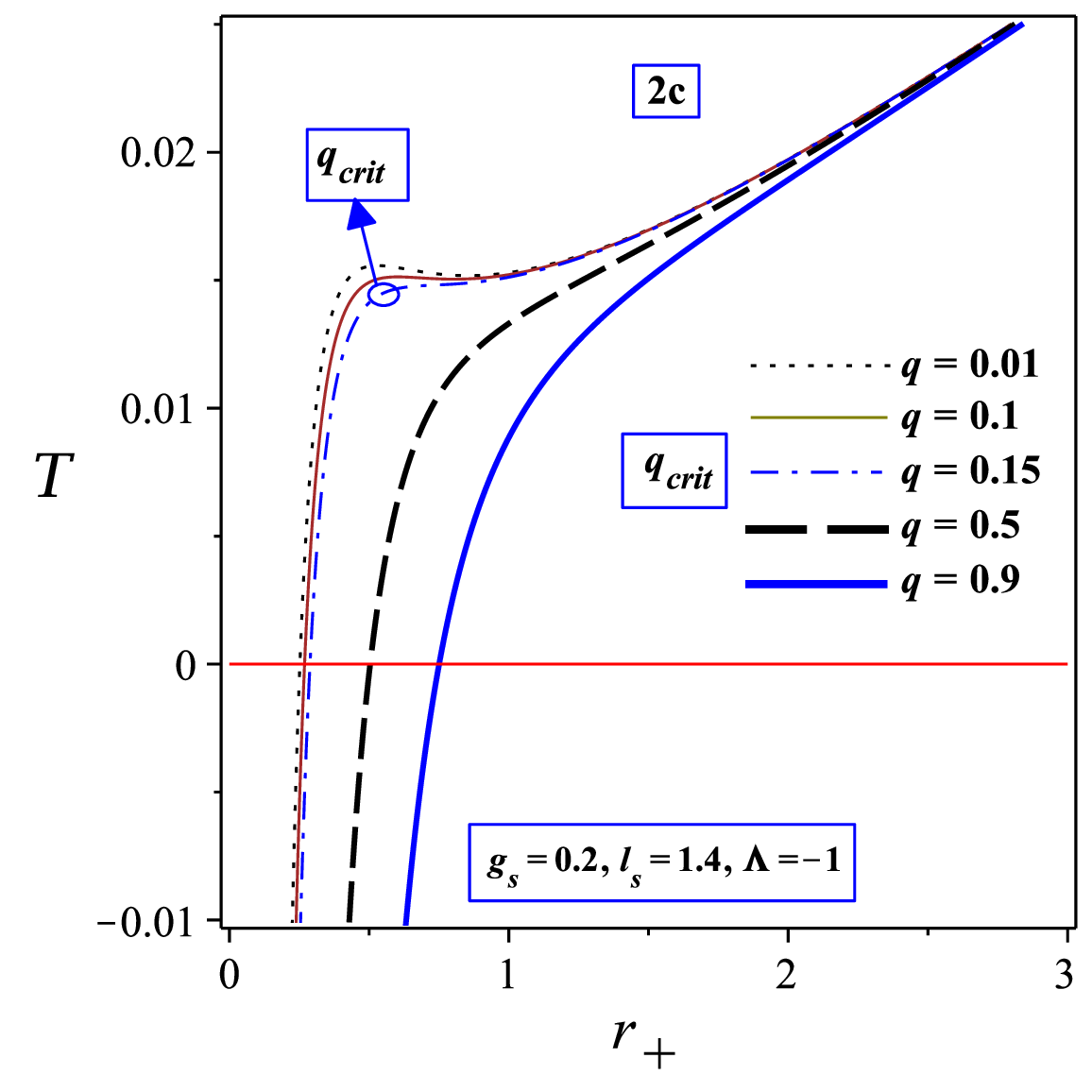}
\caption{The Hawking temperature $T$ versus $r_{+}$ for different values of
parameters $g_{s}$ (\protect\ref{fig2}a), $l_{s}$ (\protect\ref{fig2}b), and 
$q$ (\protect\ref{fig2}c).}
\label{fig2}
\end{figure}

As formulation \eqref{Temp2} indicates, the Hawking temperature $T$ is
fundamentally determined by the complete set of spacetime parameters,
specifically the electric charge ($q$), the cosmological constant ($\Lambda$%
), and the Letelier--Alencar cloud of strings parameters ($g_{s}$ and $l_{s}$%
). To systematically evaluate the influence of these parameters on the
thermal profile of the black hole, we present the behavior of $T$ as a
function of the horizon radius $r_{+}$ in Fig. \ref{fig2}. The numerical
analysis reveals the existence of a unique root for the Hawking temperature (%
$r_{+_{T=0}}$), which represents the boundary of the physical thermodynamic
region ($T > 0$). The specific effects of each parameter are detailed below:

\begin{itemize}
\item[\textbf{i)}] \textbf{Influence of $g_{s}$ (Fig. \ref{fig2}a):} An
increase in the cloud of strings parameter $g_{s}$ systematically shifts the
temperature root $r_{+_{T=0}}$ toward larger horizon radii. Furthermore, for
values below a certain threshold, $g_{s} < g_{s_{\text{crit}}}$, the
temperature curve exhibits two extrema (a minimum and a maximum), indicating
potential phase transition zones. These extrema merge and subsequently
disappear as $g_{s}$ exceeds the critical value $g_{s_{\text{crit}}}$.

\item[\textbf{ii)}] \textbf{Influence of $l_{s}$ (Fig. \ref{fig2}b):} The
thermodynamic response to the parameter $l_{s}$ qualitatively mirrors that
of $g_{s}$. Elevating $l_{s}$ increases the value of the zero-temperature
radius. The dual extrema persist only within the regime $l_{s} < l_{s_{\text{%
crit}}}$; for $l_{s} > l_{s_{\text{crit}}}$, the temperature becomes a
monotonically increasing function of $r_{+}$.

\item[\textbf{iii)}] \textbf{Influence of the electric charge $q$ (Fig. \ref%
{fig2}c):} Increasing the charge $q$ pushes the root $r_{+_{T=0}}$ to larger
values, thereby restricting the physically allowed thermodynamic domain ($T
> 0$). Similar to the previous parameters, the temperature profile
transitions from possessing two extrema for sub-critical charges ($q < q_{%
\text{crit}}$) to a strictly monotonic growth with respect to $r_{+}$ for
super-critical charges ($q > q_{\text{crit}}$).
\end{itemize}

In summary, our analysis confirms that stable, larger black holes reside
within the physical domain characterized by a positive Hawking temperature ($%
T > 0$). The parameters $g_{s}$, $l_{s}$, and $q$ act as critical regulators
of both the extent of the physical thermodynamic region and the existence of
extrema. Specifically, increasing any of these three parameters leads to a
contraction of the black hole's physically viable region.

The entropy of these black holes can be obtained by using the area law, in
the following form 
\begin{equation}
S=\frac{A}{4}=\pi r_{+}^{2},  \label{entropy}
\end{equation}%
where $A=\left. \int_{0}^{2\pi }\int_{0}^{\pi }\sqrt{g_{\theta \theta
}g_{\varphi \varphi }}\right\vert _{r=r_{+}}=4\pi \left. r^{2}\right\vert
_{r=r_{+}}=4\pi r_{+}^{2}$\ is the horizon area.

We can get the electric charge of the solutions by employ the Gauss law in
the following form 
\begin{equation}
Q=\frac{F_{tr}}{4\pi }\int_{0}^{2\pi }\int_{0}^{\pi }\sqrt{g}d\theta
d\varphi =q,  \label{Q}
\end{equation}%
where for case $t=$ constant and $r=$constant, the determinant of metric
tensor $g$ is $r^{4}\sin ^{2}\theta $.

In order to obtain electric potential, we can calculate it on the horizon
with respect to a reference which leads to 
\begin{equation}
\Phi =A_{\mu }\chi ^{\mu }\left\vert _{r\rightarrow \infty }\right. -A_{\mu
}\chi ^{\mu }\left\vert _{r\rightarrow r_{+}}\right. =\frac{q}{r_{+}}.
\label{U}
\end{equation}

Applying Ashtekar-Magnon-Das (AMD) approach, we find the total mass of these
black holes in the following form%
\begin{equation}
M=m_{0}=\frac{r_{+}}{2}+\frac{q^{2}}{2r_{+}}-\frac{\Lambda r_{+}^{3}}{6}+%
\frac{g_{s}^{2}l_{s}^{2}}{2r_{+}}\mathfrak{F}_{1_{+}},  \label{MM}
\end{equation}%
where $m_{0}$ is the geometrical mass and is given by solving $\left.
f(r)\right\vert _{r=r_{+}}=0$.

\begin{figure}[h]
\centering
\includegraphics[width=55mm]{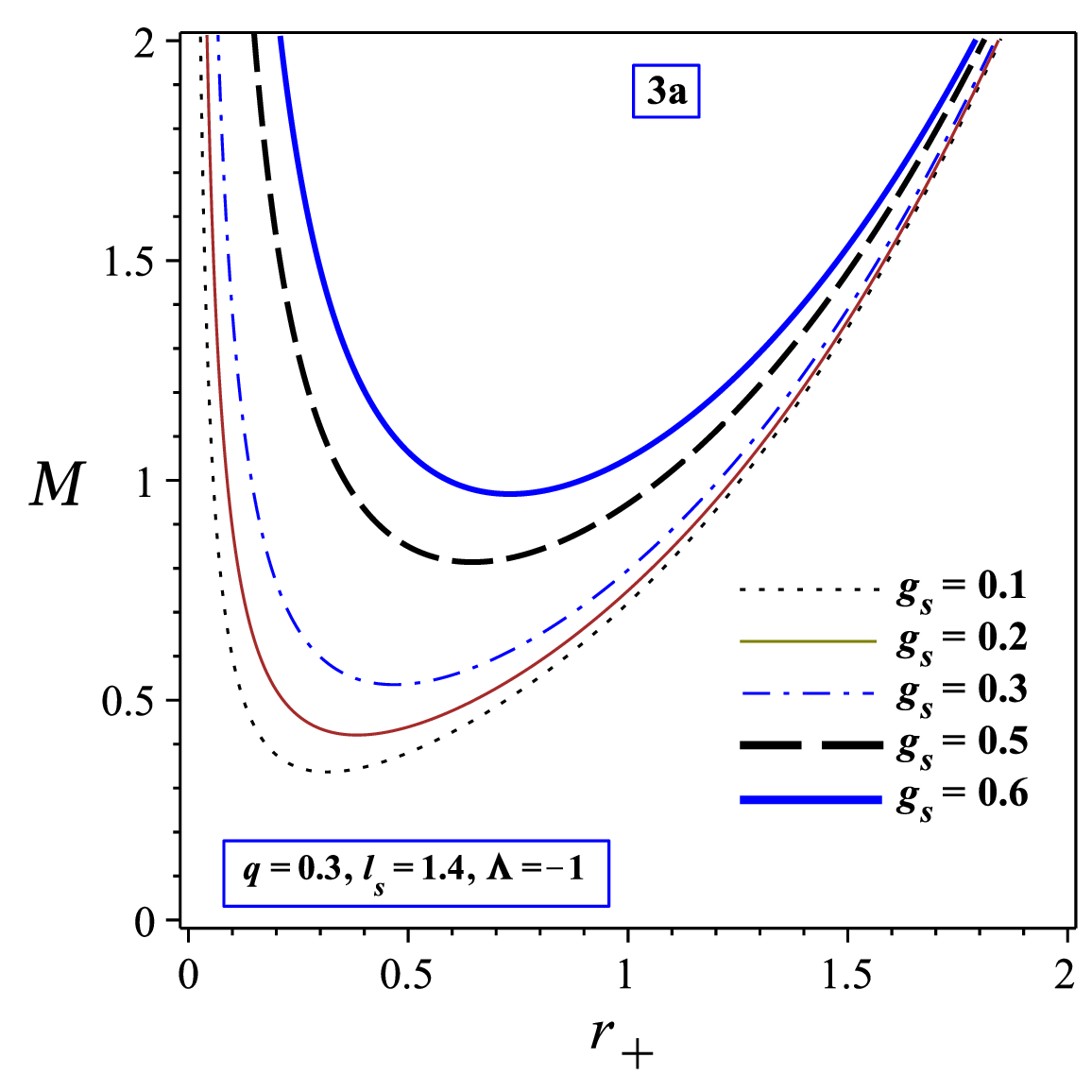} \includegraphics[width=55mm]{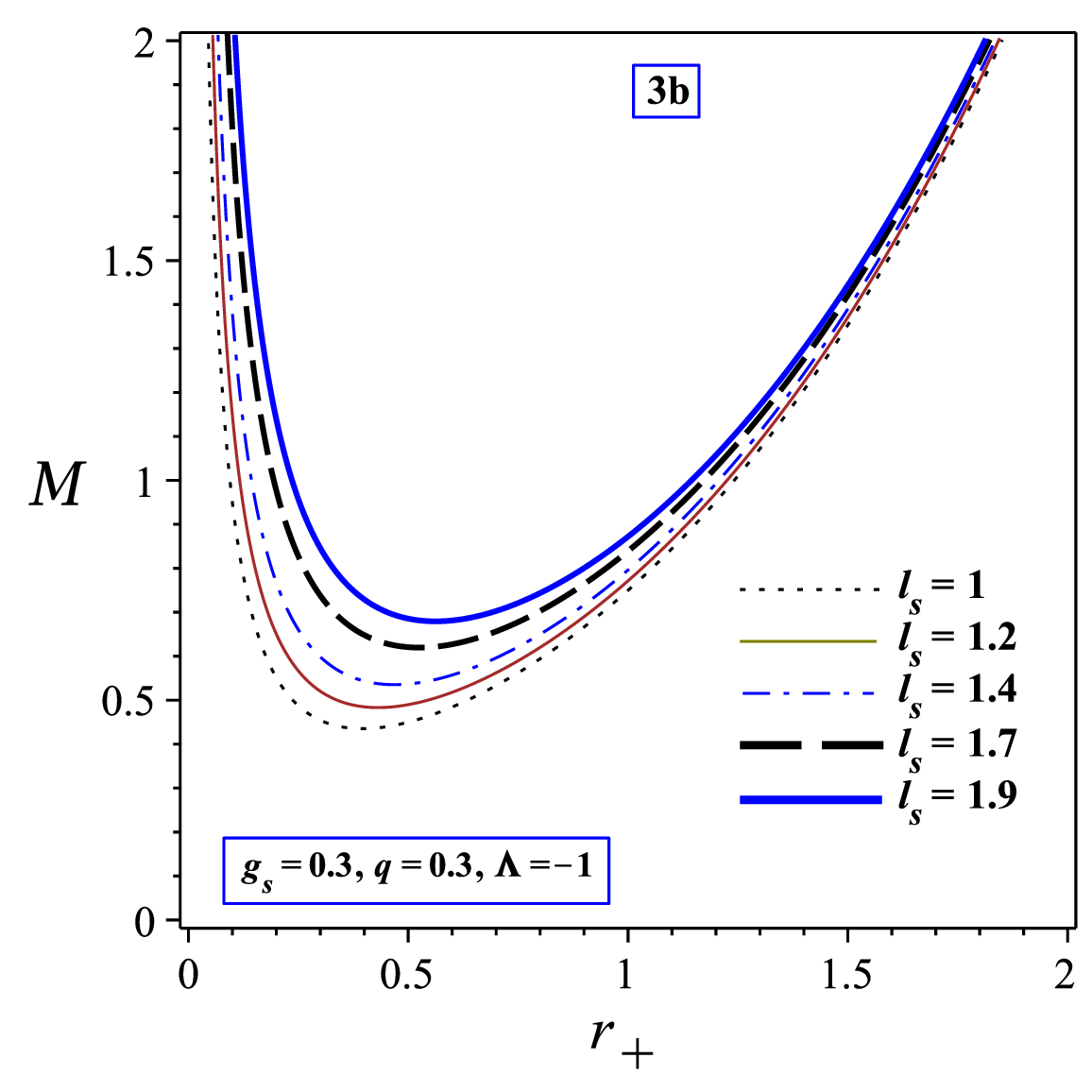} %
\includegraphics[width=55mm]{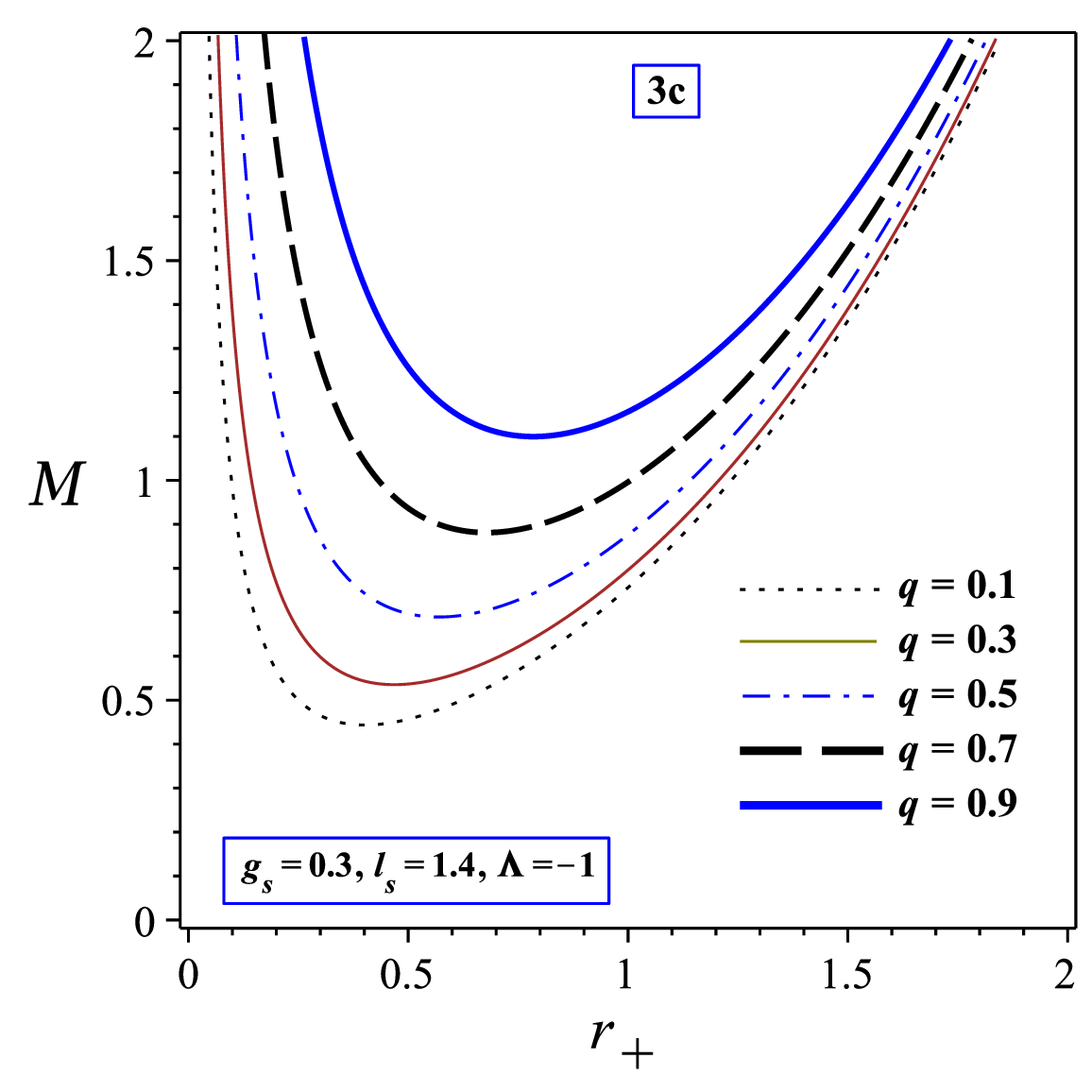}
\caption{The total mass $M$ versus $r_{+}$ for different values of
parameters $g_{s}$ (\protect\ref{fig3}a), $l_{s}$ (\protect\ref{fig3}b), and 
$q$ (\protect\ref{fig3}c).}
\label{fig3}
\end{figure}

To elucidate the influence of the system parameters on the total mass, we
illustrate the variation of $M$ as a function of the event horizon radius $%
r_{+}$ in Fig. \ref{fig3}, across three panels representing different values
of $g_{s}$, $l_{s}$, and $q$. Our analysis reveals the existence of a
minimum mass threshold, the position of which is fundamentally governed by
these parameters. Specifically, we observe that as the parameters $g_{s}$, $%
l_{s}$, and $q$ increase, the location of this mass minimum shifts
monotonically toward larger values of both $M$ and $r_{+}$.

Replacing Eqs. (\ref{entropy}) and (\ref{Q}) within Eq. (\ref{MM}), we can
obtain $M=M(S,Q)$ in the following form 
\begin{equation}
M(S,Q)=\frac{\sqrt{\frac{S}{\pi }}}{2}+\frac{Q^{2}}{2\sqrt{\frac{S}{\pi }}}-%
\frac{\Lambda S^{3/2}}{6\pi ^{3/2}}+\frac{g_{s}^{2}l_{s}^{2}}{2\sqrt{\frac{S%
}{\pi }}}{}\mathfrak{F}_{1_{S}},  \label{MS}
\end{equation}%
where $\mathfrak{F}_{1_{S}}=\left. \mathfrak{F}_{1_{+}}\right\vert _{r_{+}=%
\sqrt{\frac{S}{\pi }}}=_{2}F_{1}\left( \left[ \frac{-1}{2},\frac{-1}{4}%
\right] ,\left[ \frac{3}{4}\right] ,-\frac{S^{2}}{\pi ^{2}l_{s}^{4}}\right) $%
. Calculating derivatives of mass ($M$) with respect to entropy ($S$) and
charge ($Q$) yields temperature ($T$) and electric potential ($\Phi $),
respectively. Although the parameters of Letelier$-$Alencar affected
thermodynamic and conserved quantities, the first law remains valid as 
\begin{equation}
dM=TdS+\Phi dQ.
\end{equation}

\subsection{Heat capacity}

In the canonical ensemble, the local thermodynamic stability of a black hole
system is primarily analyzed through its heat capacity, $C_{Q}$.
Discontinuities in $C_{Q}$ function as indicators of potential thermal phase
transitions within the system. Furthermore, the sign of the heat capacity
provides a diagnostic criterion for stability: positive values ($C_{Q} > 0$)
correspond to a locally stable state, whereas negative values ($C_{Q} < 0$)
signify thermal instability \cite%
{local1,local2,local3,local4,local5,local6,local7, local8,local9}.
Additionally, the roots of the heat capacity often signify critical points
where the system undergoes a transition between stable and unstable regimes,
commonly referred to as bound points. In this study, we derive the heat
capacity for the charged Letelier--Alencar black hole solutions within the
context of the cosmological constant. Consequently, we examine the influence
of the parameters $g_{s}$, $l_{s}$, and $q$ on the regions of local
thermodynamic stability.

The heat capacity is defined as 
\begin{equation}
C=\frac{T}{\left( \frac{\partial T}{\partial S}\right) _{g_{s},l_{s},q}}%
=T\left( \frac{\frac{\partial S}{\partial r_{+}}}{\frac{\partial T}{\partial
r_{+}}}\right) _{g_{s},l_{s},q}.
\end{equation}

By applying Eqs. (\ref{Temp2}), and (\ref{entropy}) within $C$, we can get
the heat capacity as 
\begin{equation}
C=\frac{42\pi l_{s}^{4}\left( \frac{g_{s}^{2}}{r_{+}^{2}}\left( l_{s}^{4}%
\mathfrak{F}_{1_{+}}+\frac{2r_{+}^{4}}{3}\mathfrak{F}_{2_{+}}\right)
+l_{s}^{2}\left( \Lambda r_{+}^{2}+\frac{q^{2}}{r_{+}^{2}}-1\right) \right)
r_{+}^{4}}{21l_{s}^{6}\left( \Lambda r_{+}^{2}-\frac{3q^{2}}{r_{+}^{2}}%
+1\right) r_{+}^{2}-g_{s}^{2}\left( 63l_{s}^{8}\mathfrak{F}%
_{1_{+}}+12r_{+}^{8}\mathfrak{F}_{3_{+}}\right) }.  \label{C}
\end{equation}

\begin{figure}[h]
\centering
\includegraphics[width=55mm]{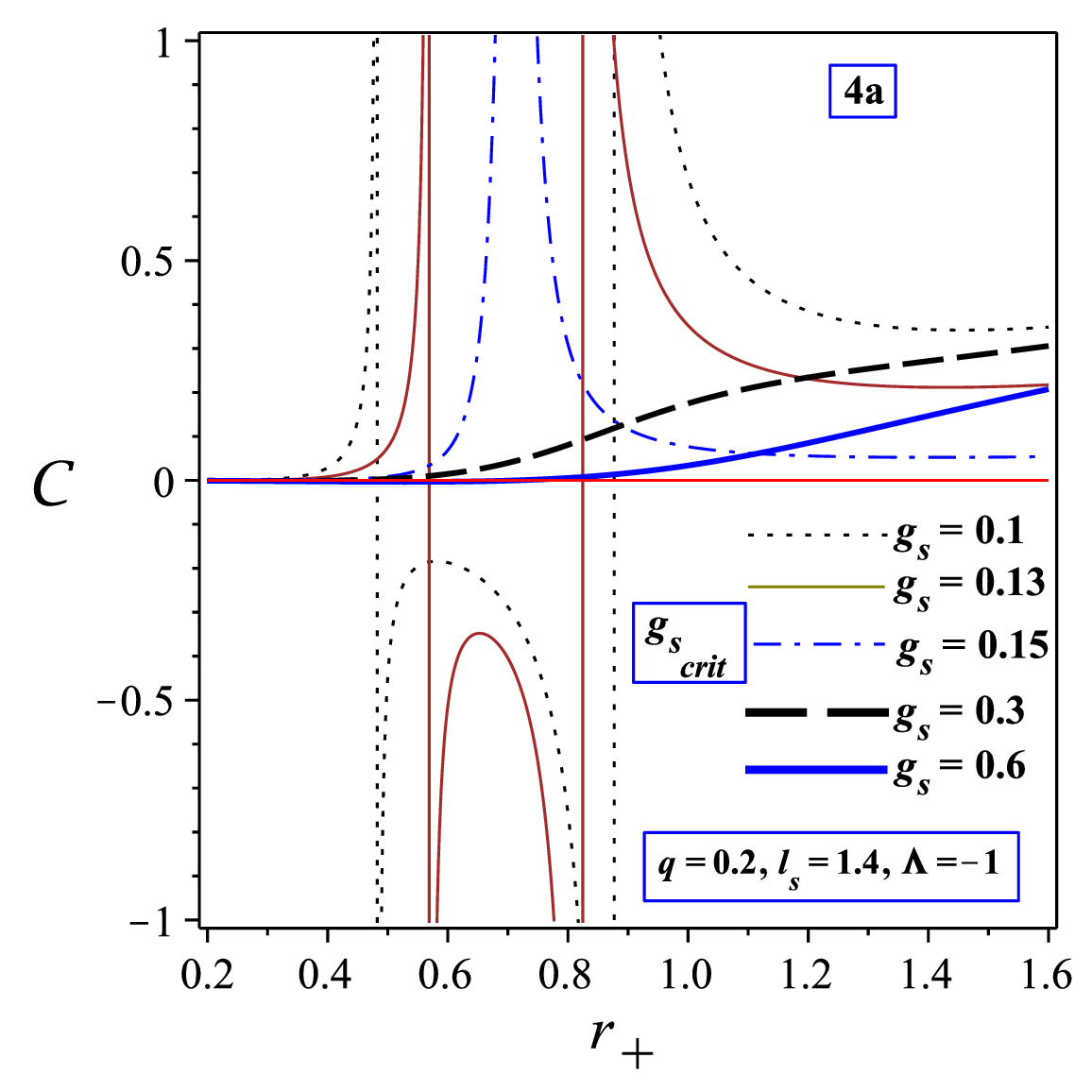} \includegraphics[width=55mm]{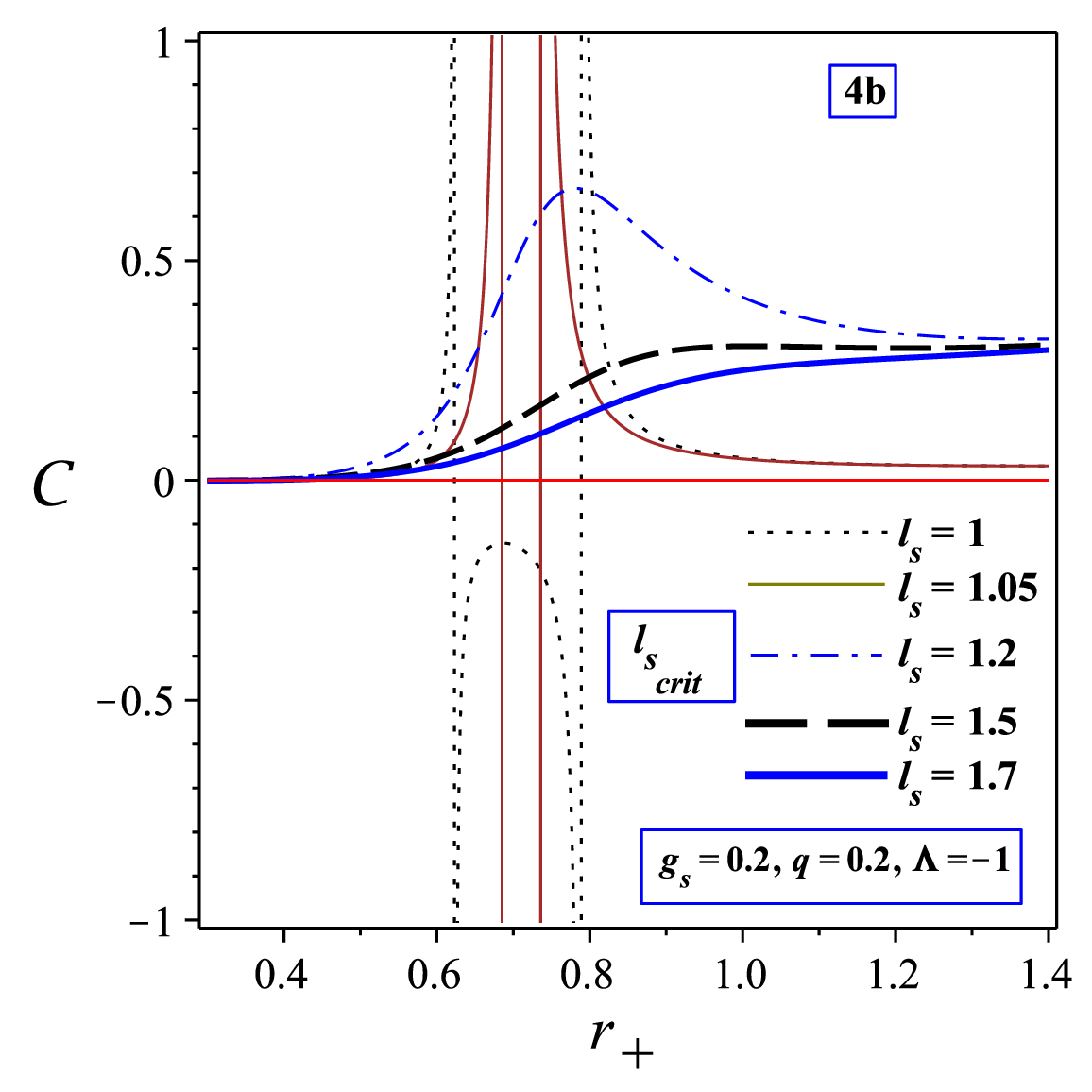} %
\includegraphics[width=55mm]{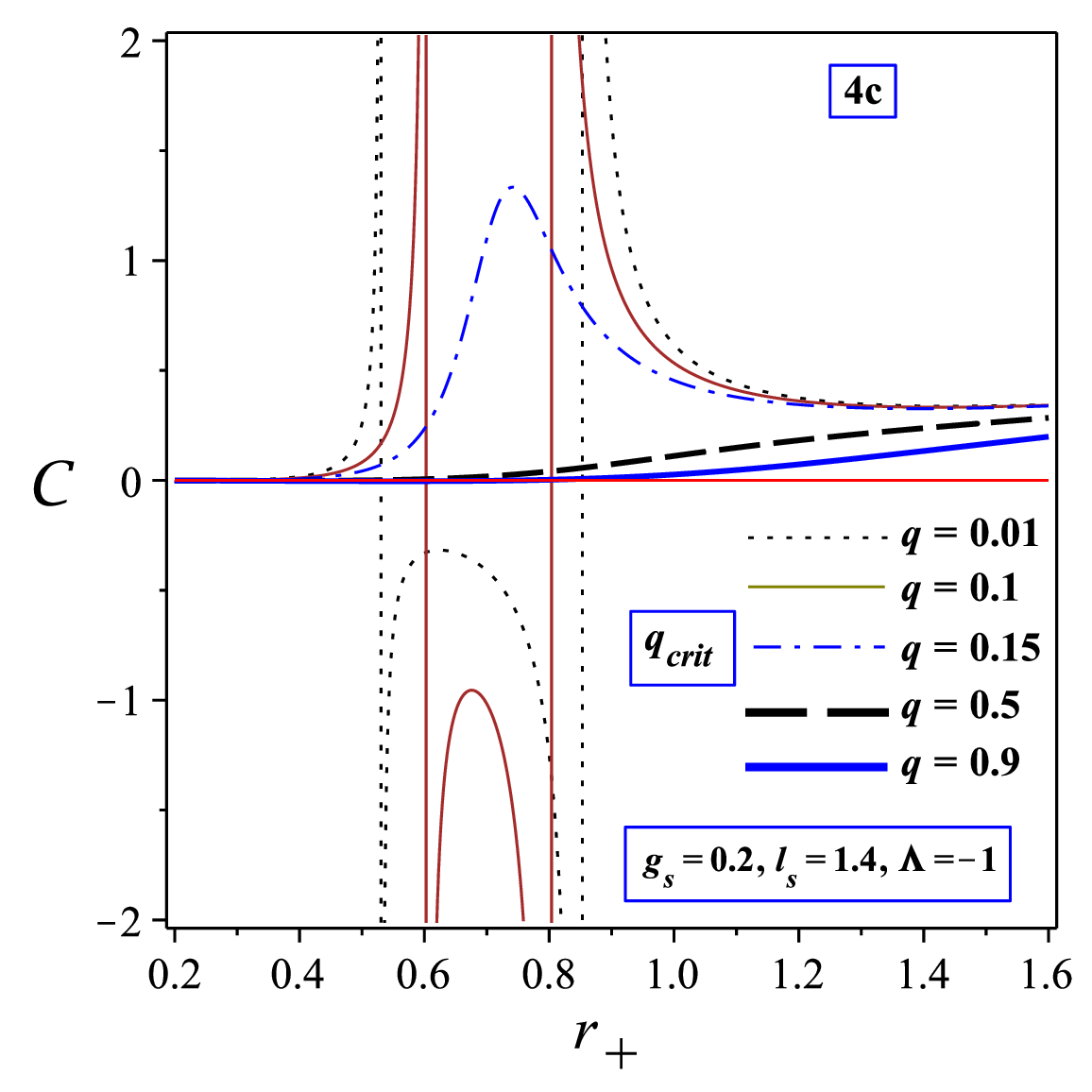}
\caption{The heat capacity $C$ versus $r_{+}$ for different values of
parameters $g_{s}$ (\protect\ref{fig4}a), $l_{s}$ (\protect\ref{fig4}b), and 
$q$ (\protect\ref{fig4}c).}
\label{fig4}
\end{figure}

By studying the heat capacity of the black hole, we can find two important
points that are related to the physical limitation and phase transition
critical points. Indeed, the root of heat capacity $\left( C=0\right) $ is
representing a border line between physical $\left( T>0\right) $\ and
non-physical $\left( T<0\right) $ black holes (which is known as the
physical limitation point). Notably, the system at this point has a change
in the sign of the heat capacity. In addition, the divergencies of the heat
capacity $\left( \left( \frac{\partial T}{\partial r_{+}}\right)
_{g_{s},l_{s},q}=0\right) $ represent phase transition critical points of
black holes. Whereas to find the exact real and positive roots and
divergence points are difficult due to complexity of relations, we plot the
heat capacity versus $r_{+}$ in Fig. \ref{fig4}. Then, we can investigate
the influence of Letelier-Alencar's \textit{cloud of strings} and the
electric charge on local stability areas, physical limitation and divergence
points. Our findings are as follows:

\textbf{i) Effect of $g_{s}$ (Fig. \ref{fig4}a):} A critical value, $%
g_{s_{crit}}$, exists. Below this value, the heat capacity ($C$) shows two
divergence points, with negative values between them, signifying a phase
transition (small/large black hole transition). Above $g_{s_{crit}}$, $C$
presents a single root, with its sign varying before and after it.

\textbf{ii) Effect of $l_{s}$ (Fig. \ref{fig4}b):} Similar to $g_{s}$, a
critical value, $l_{s_{crit}}$, is observed. Below this threshold, $C$
exhibits two divergences, indicating a phase transition between the
small/large black holes. Above $l_{s_{crit}}$, $C$ possesses a single root,
mirroring the behavior seen with $g_{s}$.

\textbf{iii) Effect of }$q$\textbf{\ (Fig. \ref{fig4}c):} For $q<q_{crit}$,
the heat capacity consistently shows two divergence points, with negative
values in between. For $q>q_{crit}$, there is one root in which the heat
capacity is negative before it and is positive after that. In addition, by
increasing $q$ leads to a reduction in the local stability region, as the
area where $C$ is negative expands.

In summary, there is a phase transition between small/large black holes when 
$g_{s}<g_{s_{crit}}$, $l_{s}<l_{s_{crit}}$, and $q<q_{crit}$. In addition,
by increasing $g_{s}$, $l_{s}$, and $q$ the positive areas of $C$ decreases
in which leads to decreasing the local stability. Importantly, large black
holes consistently satisfy the local stability conditions.

By analyzing the heat capacity of the black hole, we can identify two
essential characteristics: the boundaries of the physical domain and the
critical points of thermal phase transitions. Specifically, the roots of the
heat capacity ($C = 0$) define the interface between physical ($T > 0$) and
non-physical ($T < 0$) black hole solutions, commonly referred to as the
physical limitation point. Notably, this point coincides with a sign
inversion in $C$. Furthermore, the divergences of the heat capacity,
occurring where the condition $(\partial T / \partial r_{+})_{g_{s},l_{s},q}
= 0$ is satisfied, correspond to the phase transition critical points. Given
the analytical complexity of these expressions, which renders the
calculation of exact roots and divergence points challenging, we investigate
the thermal behavior by plotting $C$ as a function of $r_{+}$ in Fig. \ref%
{fig4}. This approach allows us to elucidate the effects of the
Letelier--Alencar \textit{cloud of strings} parameters and the electric
charge on local stability, physical boundaries, and divergence points. Our
key findings are as follows:

\begin{itemize}
\item[\textbf{i)}] \textbf{Effect of $g_{s}$ (Fig. \ref{fig4}a):} A critical
value $g_{s_{\text{crit}}}$ exists, partitioning the thermodynamic behavior.
For $g_{s} < g_{s_{\text{crit}}}$, the heat capacity exhibits two divergence
points, with a region of negative $C$ between them, indicative of a
small/large black hole phase transition. Conversely, for $g_{s} > g_{s_{%
\text{crit}}}$, the heat capacity displays a single root, accompanied by a
sign change.

\item[\textbf{ii)}] \textbf{Effect of $l_{s}$ (Fig. \ref{fig4}b):} The
influence of $l_{s}$ follows a pattern analogous to that of $g_{s}$. Below
the threshold $l_{s_{\text{crit}}}$, the system undergoes a small/large
black hole phase transition marked by two divergence points. Above this
threshold, the behavior simplifies to a single root, mirroring the regime
observed for larger values of $g_{s}$.

\item[\textbf{iii)}] \textbf{Effect of $q$ (Fig. \ref{fig4}c):} For $q < q_{%
\text{crit}}$, the heat capacity consistently features two divergence points
with an intervening region of negative values. In regime ($q > q_{\text{crit}%
}$), the system exhibits a single root, where $C$ transitions from negative
to positive. Moreover, increasing the charge $q$ constricts the local
stability domain, as the region where $C < 0$ expands.
\end{itemize}

In summary, a small/large black hole phase transition occurs under the
conditions $g_{s} < g_{s_{\text{crit}}}$, $l_{s} < l_{s_{\text{crit}}}$, and 
$q < q_{\text{crit}}$. Furthermore, increasing $g_{s}$, $l_{s}$, and $q$
reduces the positive heat capacity regions, thereby diminishing the
parameter space for local thermodynamic stability. Crucially, large black
hole solutions consistently satisfy the criteria for local stability across
these parameter ranges.

\begin{figure}[h]
\centering
\includegraphics[width=70mm]{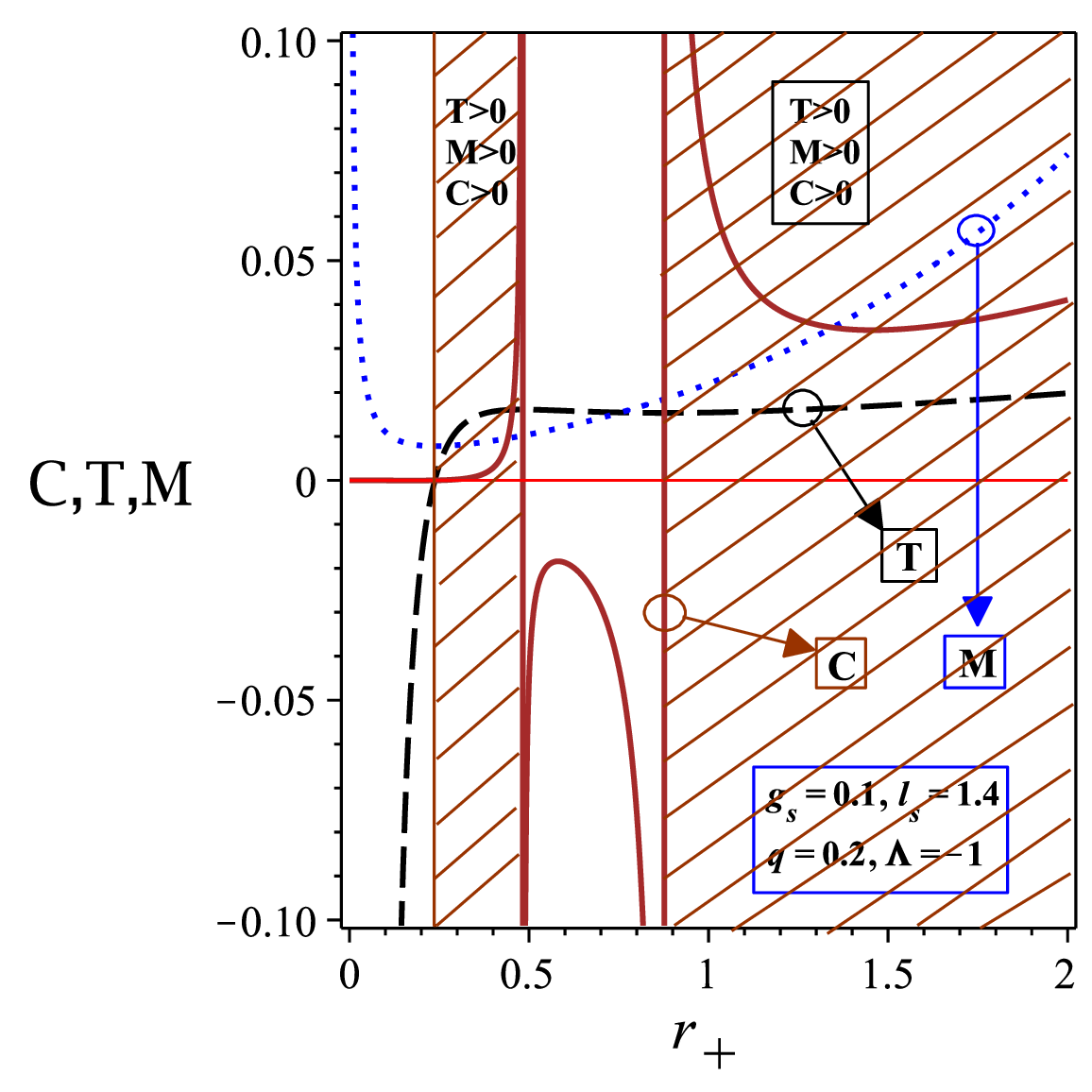}
\caption{The total mass $M$, the heat capacity $C$ and temperature $T$
versus $r_{+}$ for small value of $g_{s}$, $l_{s}$ and $q$.}
\label{fig5}
\end{figure}

We now examine the parametric dependence of the black hole solutions, as
illustrated in Fig. \ref{fig5}. Notably, for sufficiently small values of $%
g_{s}$, $l_{s}$, and $q$, the phase transition between small and large black
holes is suppressed (see Fig. \ref{fig3}). The parameter space corresponding
to this phase transition is characterized by positive values for the mass ($%
M $), Hawking temperature ($T$), and heat capacity ($C$) simultaneously.
Consequently, this region fulfills the essential requirements for local
thermodynamic stability.

\subsection{Gibbs potential}

The foundational framework for assessing the global thermodynamic stability
of black holes was pioneered by Hawking and Page \cite{Hawking:1983}. Within
the context of the grand canonical ensemble, global stability is typically
evaluated by analyzing the Gibbs free energy ($G$); a black hole
configuration is considered globally stable if it exhibits a negative Gibbs
potential \cite{Gibbs1,Gibbs2,Gibbs3,Gibbs4,Gibbs5}. In this section, we
apply this formalism to investigate the global stability of charged
Letelier--Alencar black holes in the presence of a cosmological constant.

The Gibbs potential is defined as 
\begin{equation}
G=M-TS-\Phi Q.  \label{G}
\end{equation}

Using Eqs. (\ref{Temp2}), (\ref{entropy}), (\ref{Q}), (\ref{U}), and (\ref%
{MM}) within Eq. (\ref{G}), we find the Gibbs potential in the following
form 
\begin{equation}
G=\frac{\Lambda r_{+}^{3}}{12}+\frac{r_{+}}{4}-\frac{q^{2}}{4r_{+}}%
+g_{s}^{2}\left( \frac{3l_{s}^{2}\mathfrak{F}_{1_{+}}}{4r_{+}}+\frac{%
r_{+}^{3}\mathfrak{F}_{2_{+}}}{6l_{s}^{2}}\right) ,  \label{GG}
\end{equation}%
where depends on all of parameters such as the cosmological constant, $g_{s}$%
, $l_{s}$ and $q$. It is notable that according to the equation (\ref{Q}), $%
q=Q$.

\begin{figure}[h]
\centering
\includegraphics[width=55mm]{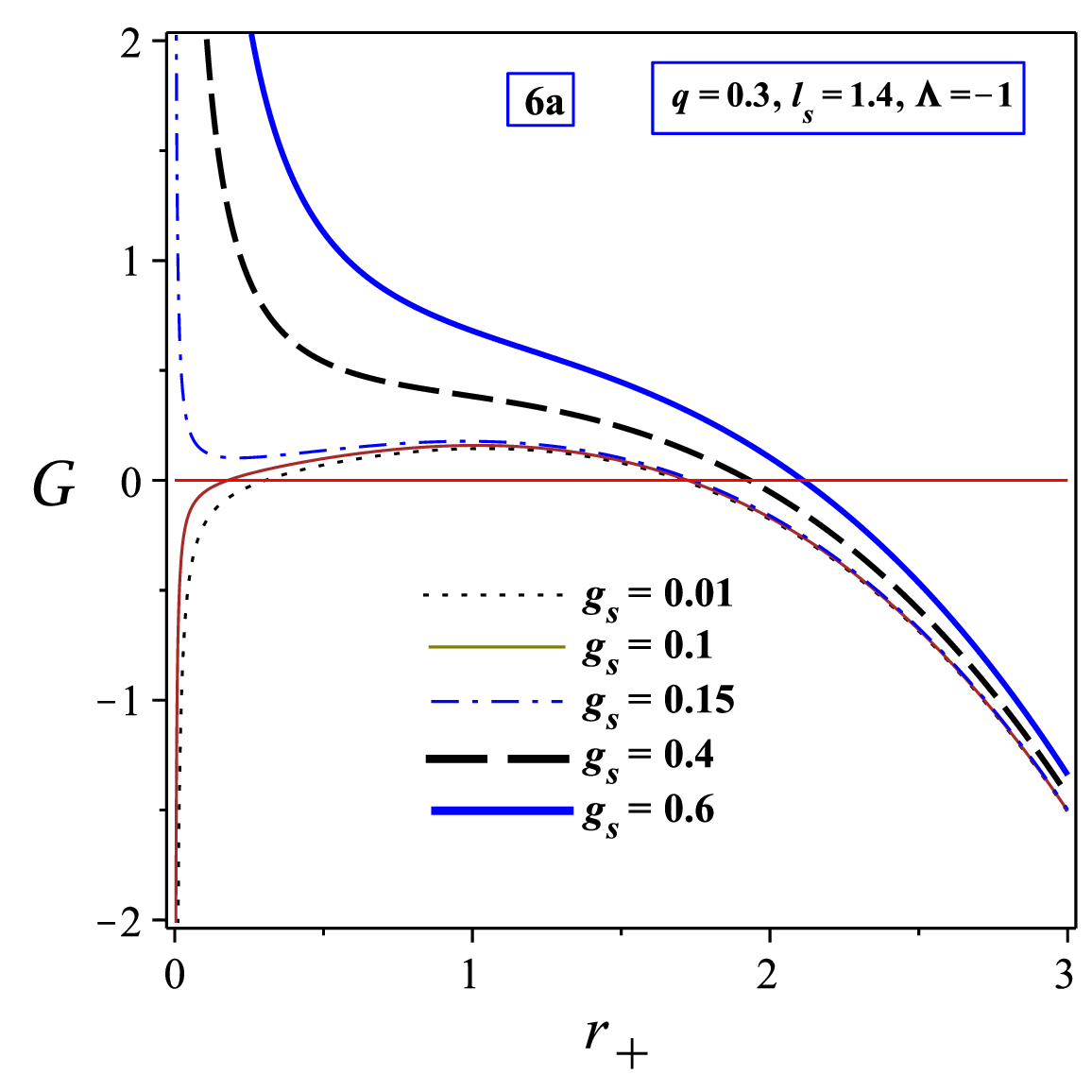} \includegraphics[width=55mm]{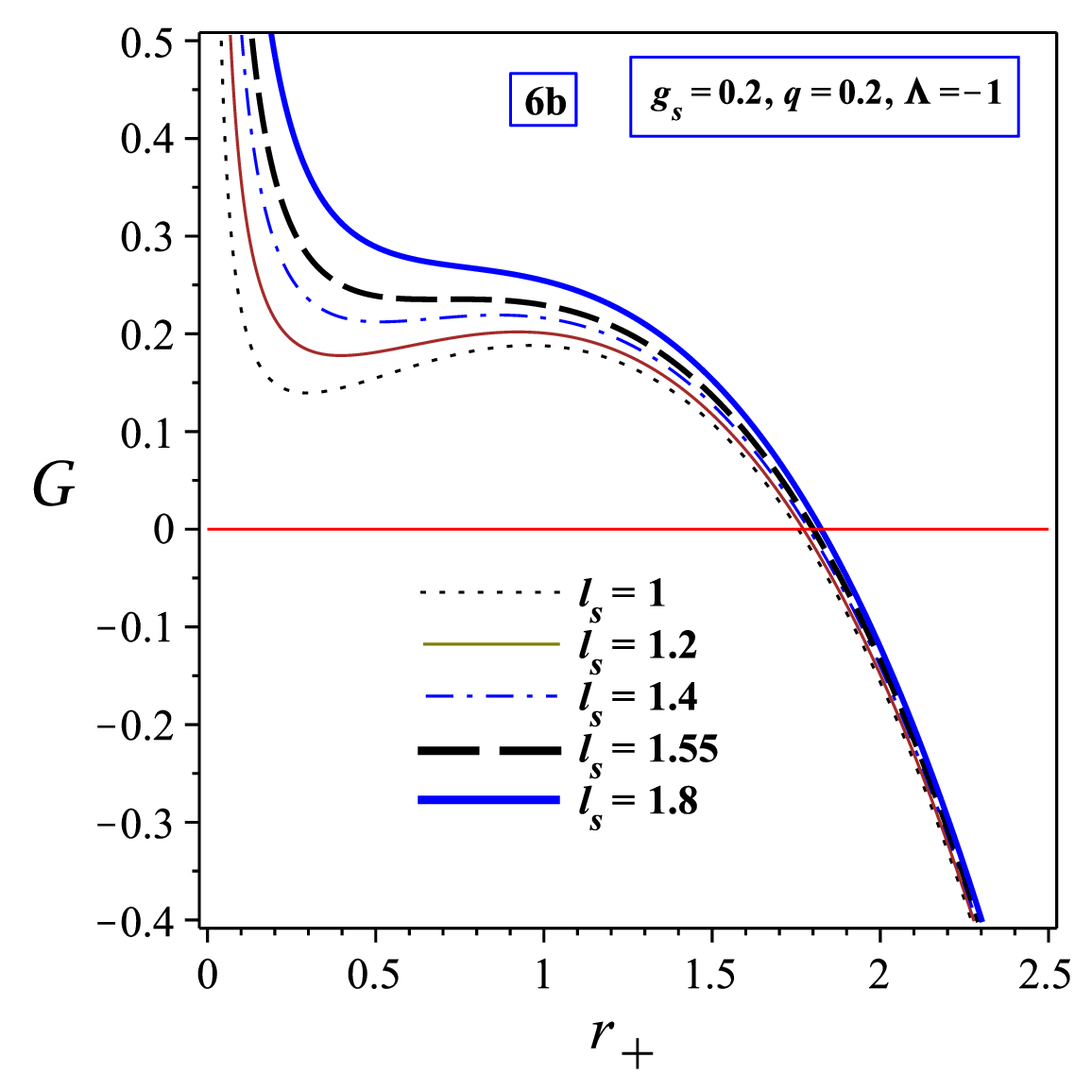} %
\includegraphics[width=55mm]{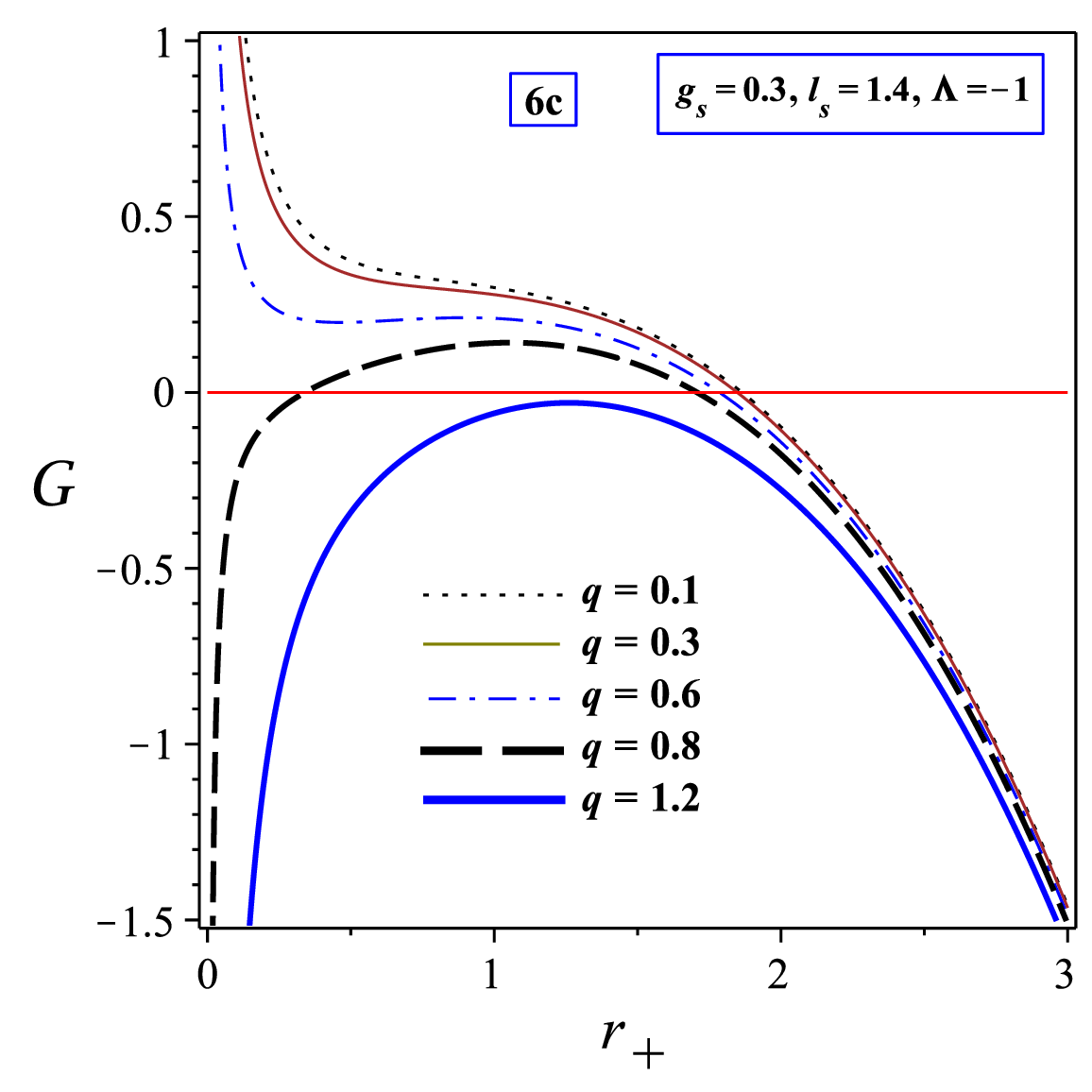}
\caption{The total mass $F$ versus $r_{+}$ for different values of
parameters $g_{s}$ (\protect\ref{fig6}a), $l_{s}$ (\protect\ref{fig6}b), and 
$q$ (\protect\ref{fig6}c).}
\label{fig6}
\end{figure}

Figure \ref{fig6} illustrates the impact of the parameters $g_{s}$, $l_{s}$,
and $q$ on the global stability regions of the black hole. Our analysis
leads to the following observations:

\begin{itemize}
\item[\textbf{i)}] \textbf{Effect of $g_{s}$ (Fig. \ref{fig6}a):} A critical
value $g_{s_{\text{crit}}}$ (represented by the blue dashed-dotted line)
delineates two distinct thermodynamic regimes. For $g_{s}<g_{s_{\text{crit}%
}} $, the Gibbs potential $G$ possesses two roots ($r_{+_{1}}$ and $r_{+_{2}}
$). In the intermediate range $r_{+_{1}} < r_{+} < r_{+_{2}}$, the black
hole fails to satisfy the global stability condition. Conversely, for $r_{+}
< r_{+_{1}}$ and $r_{+} > r_{+_{2}}$, the condition $G < 0$ is satisfied,
indicating globally stable configurations. For $g_{s} > g_{s_{\text{crit}}}$%
, the Gibbs potential exhibits a single root, with $G$ remaining negative
for radii larger than this root; consequently, large black holes remain
globally stable. Notably, the domain of global stability contracts as $g_{s}$
increases.

\item[\textbf{ii)}] \textbf{Effect of $l_{s}$ (Fig. \ref{fig6}b):} The Gibbs
potential displays a single root. For radii below this root, $G > 0$,
implying that small black holes fail to satisfy the global stability
criterion. However, for radii beyond the root, $G < 0$, ensuring global
stability for large black holes. An increase in $l_{s}$ shifts the root
toward larger values of $r_{+}$, thereby reducing the global stability
region.

\item[\textbf{iii)}] \textbf{Effect of $q$ (Fig. \ref{fig6}c):} The system
exhibits two critical values, $q_{\text{crit}_{1}}$ and $q_{\text{crit}_{2}}$%
. For $q < q_{\text{crit}_{1}}$, the Gibbs potential is negative across the
entire domain, implying global stability for all black holes. In the range $%
q_{\text{crit}_{1}} < q < q_{\text{crit}_{2}}$, two roots emerge; global
stability is maintained in the intervals $r_{+} < r_{+_{1}}$ and $r_{+} >
r_{+_{2}}$. For $q > q_{\text{crit}_{2}}$, the potential admits a single
root, and global stability is restricted to large black holes ($r_{+} >
r_{+_{\text{root}}}$). Consistent with the other parameters, the global
stability region diminishes as $q$ increases.
\end{itemize}

In summary, increasing $g_{s}$, $l_{s}$, and $q$ leads to a reduction in the
globally stable parameter space, as the regions where $G > 0$ expand.
Furthermore, large black holes consistently satisfy the condition for global
stability, which is in excellent agreement with the behavior previously
observed for local stability.

\section{Constraining the Black Hole Parameters from QPO Observations}

\label{bi}

In this section, we constrain the parameters of the black hole solution
using quasi-periodic oscillation (QPO) observations. Since astrophysical
black holes are effectively asymptotically flat, we set the cosmological
constant to zero, i.e., $\Lambda=0$.

The motion of a test particle in the black hole spacetime is described by
the Lagrangian

\begin{equation}
L_{p}=\frac{1}{2}mg_{\mu \nu }\dot{x}^{\mu }\dot{x}^{\nu },
\end{equation}

where $m$ denotes the mass of the particle and an overdot represents
differentiation with respect to the proper time $\tau$. The particle
trajectory is given by $x^\mu(\tau)$, while its four-velocity is defined as

\begin{equation}
u^{\mu }=\frac{dx^{\mu }}{d\tau }.
\end{equation}

Owing to the static and spherically symmetric nature of the spacetime, two
Killing vectors exist,

\begin{equation}
\xi ^{\mu }=(1,0,0,0),~~~\&~~~\eta ^{\mu }=(0,0,0,1),
\end{equation}

which correspond to time-translation and rotational symmetries,
respectively. These symmetries lead to two conserved quantities: the
specific energy $\mathcal{E}$ and the specific angular momentum $\mathcal{L}$
of the test particle,

\begin{align}
\mathcal{E}&=-g_{tt}\dot{t},  \notag \\
\mathcal{L}&=g_{\phi\phi}\dot{\phi}.  \label{energy}
\end{align}

The equations of motion are further constrained by the normalization
condition of the four-velocity,

\begin{equation}
g_{\mu \nu }u^{\mu }u^{\nu }=\delta ,  \label{normal}
\end{equation}

where $\delta=0$ corresponds to null geodesics, while $\delta=-1$ and $%
\delta=+1$ describe timelike and spacelike geodesics, respectively. Since
QPOs originate from matter orbiting the black hole, we restrict our analysis
to massive particles following timelike geodesics ($\delta=-1$).

For a static and spherically symmetric spacetime, the motion may be confined
to the equatorial plane by choosing

\begin{equation}
\theta =\frac{\pi }{2},~~~\&~~~\dot{\theta}=0.
\end{equation}

Substituting Eq.~(\ref{energy}) into the normalization condition (\ref%
{normal}), the radial equation of motion takes the form

\begin{equation}
\dot{r}^{\,2}=\mathcal{E}^{2}-V_{\mathrm{eff}}(r),  \label{radial}
\end{equation}

where the effective potential is given by

\begin{equation}
V_{\mathrm{eff}}(r)=-g_{tt}(r)\left( 1+\frac{\mathcal{L}^{2}}{r^{2}}\right) .
\end{equation}

Circular orbits are characterized by constant radial coordinate and
therefore satisfy the conditions

\begin{equation}
\dot{r}=0,~~~\&~~~\ddot{r}=0,
\end{equation}

or, equivalently,

\begin{equation}
V_{\mathrm{eff}}(r)=\mathcal{E}^{2},~~~\&~~~\frac{dV_{\mathrm{eff}}}{dr}=0.
\label{const}
\end{equation}

These conditions determine the specific energy and angular momentum of
particles moving along circular geodesics and serve as the starting point
for calculating the orbital and epicyclic frequencies employed in the QPO
analysis.

\begin{figure*}[!t]
\centering
\includegraphics[width=0.32\linewidth]{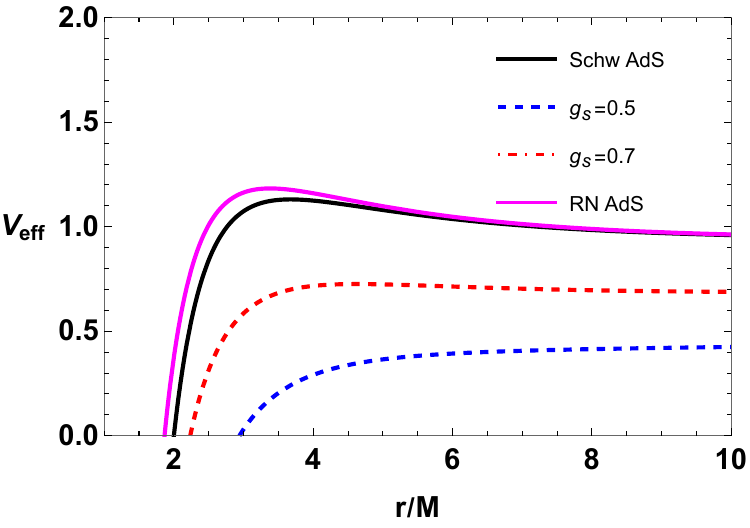} 
\includegraphics[width=0.32\linewidth]{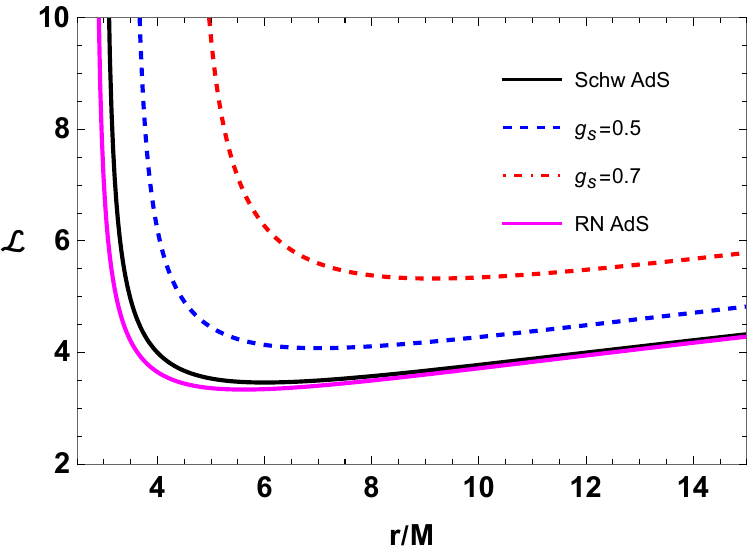} %
\includegraphics[width=0.32\linewidth]{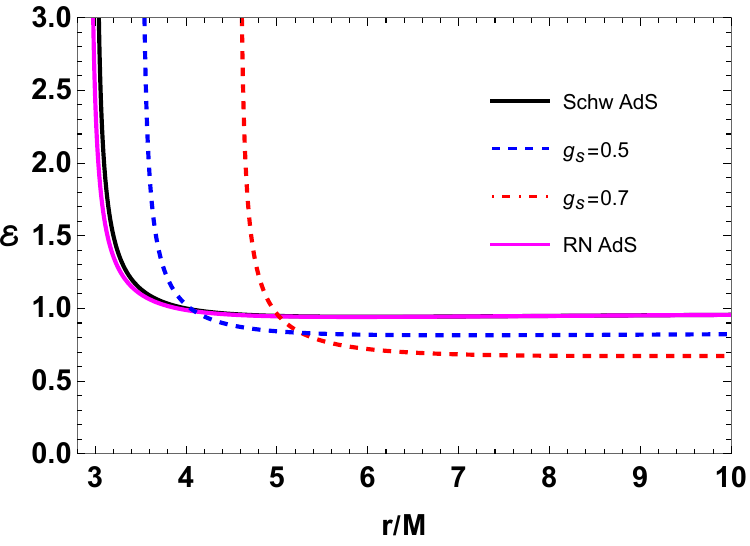} \vspace{-0.3cm}
\caption{The radial dependence of effective potential, specific angular
momentum and energy for circular orbits for different values of coupling
parameter $g_s$. Here, we have considered $q=0.5,l_s=0.5,\Lambda=0,m_{0}=1~ 
\text{and} ~L=20$}
\label{fig1b}
\end{figure*}

Next, using the expressions Eq. (\ref{const}) , we derive expressions for
the specific angular momentum and the specific energy as : 
\begin{eqnarray}
\mathcal{L} &=&-\frac{r^{2}\left( 3g_{s}^{2}l_{s}^{2}\,\mathfrak{F}%
_{1}+3g_{s}^{2}l_{s}^{2}\sqrt{\frac{r^{4}}{l_{s}^{4}}+1}-6\text{m}%
_{0}r+6q^{2}+2\Lambda r^{4}\right) }{9g_{s}^{2}l_{s}^{2}\,\mathfrak{F}%
_{1}+3g_{s}^{2}l_{s}^{2}\sqrt{\frac{r^{4}}{l_{s}^{4}}+1}-18\text{m}%
_{0}r+12q^{2}+6r^{2}}, \\
&&  \notag \\
\mathcal{E} &=&\frac{2\left( 3g_{s}^{2}l_{s}^{2}\,\mathfrak{F}_{1}-r\left( 6%
\text{m}_{0}+\Lambda r^{3}-3r\right) +3q^{2}\right) {}^{2}}{9r^{2}\left(
3g_{s}^{2}l_{s}^{2}\mathfrak{F}_{1}+g_{s}^{2}l_{s}^{2}\sqrt{\frac{r^{4}}{%
l_{s}^{4}}+1}-6\text{m}_{0}r+4q^{2}+2r^{2}\right) }.
\end{eqnarray}

Figure~\ref{fig1b} illustrates the radial profiles of the effective
potential $V_{\mathrm{eff}}$, the specific angular momentum $\mathcal{L}$,
and the specific energy $\mathcal{E}$ required for circular geodesic motion
for different values of the coupling parameter $g_{s}$. For comparison, the
Schwarzschild--AdS ($g_{s}=0$, and $q=0$) and Reissner--Nordstr\"{o}m--AdS ($%
g_{s}=0$) black holes are also shown. Throughout the analysis, we fix the
parameters $q=0.5$, $l_{s}=0.5$, $\Lambda =0$, $m_{0}=1$, and $L=20$.

The left panel depicts the effective potential as a function of the radial
coordinate. $V_{\mathrm{eff}}$ exhibits a peak which represents the
potential barrier that separates stable and unstable orbital regions. The
location and height of this maximum depend on the coupling parameter $g_s$.
As $g_s$ increases, the peak of the effective potential becomes lower. The
effective potentials gradually converge towards the RN profile as the $g_s$
value approaches $0$.

The middle panel depicts the specific angular momentum required to sustain
circular orbits. In each case, $\mathcal{L}$ exhibits a minimum that
corresponds to the location of the innermost stable circular orbit (ISCO).
As the coupling parameter decreases, both the minimum value of $\mathcal{L}$
and the radius at which it occurs shift to larger values. Consequently,
stronger coupling effects require orbiting matter to possess larger angular
momentum in order to maintain stable circular motion.

The right panel shows the corresponding specific energy of circular orbits.
Similar to the angular momentum, the energy profile possesses a minimum
which is the transition point from unstable to stable circular orbits. The
coupling decreases the energy required for circular motion.

To investigate the oscillatory motion of test particles around the black
hole, we consider small perturbations about a stable circular orbit. These
perturbations give rise to the radial and vertical (latitudinal) epicyclic
oscillations. Let the equilibrium circular orbit be located at $%
(r_{0},\theta _{0})$, and introduce small deviations according to 
\begin{equation}
r=r_{0}+\delta r,~~~\&~~~\theta =\theta _{0}+\delta \theta ,
\end{equation}%
where $|\delta r|\ll r_{0}$ and $|\delta \theta |\ll 1$.

Expanding the effective potential about the equilibrium orbit up to second
order yields 
\begin{align}
V_{\mathrm{eff}}(r,\theta )& =V_{\mathrm{eff}}(r_{0},\theta _{0})+\delta
r\left. \frac{\partial V_{\mathrm{eff}}}{\partial r}\right\vert _{0}+\delta
\theta \left. \frac{\partial V_{\mathrm{eff}}}{\partial \theta }\right\vert
_{0}+\frac{1}{2}\delta r^{2}\left. \frac{\partial ^{2}V_{\mathrm{eff}}}{%
\partial r^{2}}\right\vert _{0}  \notag \\
&  \notag \\
& \quad +\frac{1}{2}\delta \theta ^{2}\left. \frac{\partial ^{2}V_{\mathrm{%
eff}}}{\partial \theta ^{2}}\right\vert _{0}+\delta r\,\delta \theta \left. 
\frac{\partial ^{2}V_{\mathrm{eff}}}{\partial r\partial \theta }\right\vert
_{0}+\mathcal{O}(\delta ^{3}),  \label{expansion}
\end{align}%
where the subscript \textquotedblleft 0\textquotedblright\ denotes
evaluation at $(r_{0},\theta _{0})$.

Since the particle follows a circular geodesic, the first derivatives of the
effective potential vanish, 
\begin{equation}
\left. \frac{\partial V_{\mathrm{eff}}}{\partial r}\right\vert _{0}=\left. 
\frac{\partial V_{\mathrm{eff}}}{\partial \theta }\right\vert _{0}=0,
\end{equation}%
and the leading contribution arises from the quadratic terms. Consequently,
the perturbation equations reduce to two independent harmonic oscillator
equations, 
\begin{equation}
\frac{d^{2}\delta r}{dt^{2}}+\Omega _{r}^{2}\,\delta r=0,~~~\&~~~\frac{%
d^{2}\delta \theta }{dt^{2}}+\Omega _{\theta }^{2}\,\delta \theta =0,
\end{equation}%
where $\Omega _{r}$ and $\Omega _{\theta }$ denote the radial and vertical
epicyclic angular frequencies, respectively, as measured by a distant
observer \cite{Bambi2017book}. These frequencies are given by 
\begin{equation}
\Omega _{r}^{2}=-\frac{1}{2g_{rr}\dot{t}^{\,2}}\left. \frac{\partial ^{2}V_{%
\mathrm{eff}}}{\partial r^{2}}\right\vert _{\theta =\pi /2},  \label{Omegar2}
\end{equation}%
and 
\begin{equation}
\Omega _{\theta }^{2}=-\frac{1}{2g_{\theta \theta }\dot{t}^{\,2}}\left. 
\frac{\partial ^{2}V_{\mathrm{eff}}}{\partial \theta ^{2}}\right\vert
_{\theta =\pi /2}.  \label{Omegatheta}
\end{equation}

For the present black hole spacetime, the radial epicyclic frequency assumes
a lengthy analytical form 
\begin{equation}
\Omega _{r}=\frac{1}{\sqrt{6}\,r^{3}}\sqrt{\mathcal{R}(r)},  \label{Omegar3}
\end{equation}%
where 
\begin{align}
\mathcal{R}(r)=& -9g_{s}^{4}l_{s}^{4}\,{}\mathfrak{F}_{1}+6g_{s}^{4}\left(
l_{s}^{4}+r^{4}\right) +3g_{s}^{2}\left( 1+\frac{r^{4}}{l_{s}^{4}}\right) 
\mathfrak{F}_{4}{}\Bigg[g_{s}^{2}(3l_{s}^{4}+5r^{4})+l_{s}^{2}\sqrt{1+\frac{%
r^{4}}{l_{s}^{4}}}\left( (-12m_{0}+5\Lambda r^{3}+r)r+9q^{2}\right) \Bigg] 
\notag \\
&  \notag \\
& +\frac{g_{s}^{2}}{l_{s}^{2}\sqrt{1+\frac{r^{4}}{l_{s}^{4}}}}\Big[%
3l_{s}^{4}\left( -(6m_{0}+r)r+7q^{2}+3\Lambda r^{4}\right) +3r^{4}\left(
(r-10m_{0})r+9q^{2}\right) +7\Lambda r^{8}\Big]  \notag \\
&  \notag \\
& +36m_{0}^{2}r^{2}+6q^{2}(4\Lambda r^{3}-9m_{0})r-6m_{0}(5\Lambda
r^{5}+r^{3})+24q^{4}+8\Lambda r^{6},
\end{align}%
where $\mathfrak{F}_{4}={}_{2}F_{1}\left( \left[ 1,\frac{5}{4}\right] ,\left[
\frac{3}{4}\right] ,-\frac{r^{4}}{l_{s}^{4}}\right) $.

The vertical epicyclic frequency admits the expression 
\begin{equation}
\Omega _{\theta }^{2}=\Omega _{\phi }^{2}=\sqrt{-\frac{3g_{s}^{2}l_{s}^{2}%
\left( \,\mathfrak{F}_{1}+\sqrt{\frac{r^{4}}{l_{s}^{4}}+1}\right) -6\text{m}%
_{0}r+6q^{2}+2\Lambda r^{4}}{6r^{4}},}
\end{equation}%
which, owing to the spherical symmetry of the spacetime, is identical to the
orbital (Keplerian) angular frequency, i.e., $\Omega _{\theta }=\Omega
_{\phi }$.

Finally, the corresponding observable frequencies measured in Hertz are
obtained by restoring the physical constants, 
\begin{equation}
\nu _{i}=\frac{c^{3}}{2\pi GM}\,\Omega _{i},~~~~~~i=\phi ,r,\theta ,
\end{equation}%
where $G$ and $c$ denote the gravitational constant and the speed of light,
respectively.\newline

\subsection{QPO Models}

In this section, we investigate the twin-peak QPOs associated with the black
hole solution under consideration and compare the resulting frequency
correlations with those of the Schwarzschild and RN black holes. The
observable upper ($\nu_U$) and lower ($\nu_L$) QPO frequencies are
constructed from the orbital and epicyclic frequencies derived in the
previous subsection. To confront the theoretical predictions with
observations, we consider several widely used geodesic QPO models \cite{51},
namely the relativistic precession (RP), warped disk (WD), and epicyclic
resonance (ER) models.

\begin{itemize}
\item \textbf{Relativistic Precession (RP) model:} The RP model attributes
the observed QPOs to the fundamental frequencies of geodesic motion of
matter orbiting the compact object \cite%
{StellaVietri1998,StellaVietri1999,MorsinkStella1999}. In this framework,
the upper and lower high-frequency QPOs are identified as 
\begin{equation}
\nu _{U}=\nu _{\phi },~~~\&~~~\nu _{L}=\nu _{\phi }-\nu _{r},
\end{equation}%
where $\nu _{\phi }$ and $\nu _{r}$ denote the orbital and radial epicyclic
frequencies, respectively. The low-frequency QPO is associated with the
nodal precession frequency, $\nu _{\phi }-\nu _{\theta }$, which vanishes
for static spherically symmetric spacetimes since $\nu _{\theta }=\nu _{\phi
}$.

\item \textbf{Epicyclic Resonance (ER) models:} The ER models interpret QPOs
as arising from nonlinear resonances between the radial and vertical
epicyclic oscillations in the inner accretion disk. The frequency
identifications considered in this work are 
\begin{align}
\text{ER2:}\qquad & \nu _{U}=2\nu _{\theta }-\nu _{r},~~~\&~~~\nu _{L}=\nu
_{r}, \\
\text{ER3:}\qquad & \nu _{U}=\nu _{\theta }+\nu _{r},~~~\&~~~\nu _{L}=\nu
_{\theta }, \\
\text{ER4:}\qquad & \nu _{U}=\nu _{\theta }+\nu _{r},~~~\&~~~\nu _{L}=\nu
_{\theta }-\nu _{r}.
\end{align}

\item \textbf{Warped Disk (WD) model:} The WD model explains high-frequency
QPOs as nonlinear resonances between oscillation modes of a warped accretion
disk \cite{wd1,wd2}. In this scenario, the observed frequencies are given by 
\begin{equation}
\nu _{U}=2\nu _{\phi }-\nu _{r},~~~\&~~~\nu _{L}=2(\nu _{\phi }-\nu _{r}).
\end{equation}
\end{itemize}

Using these frequency prescriptions, we compute the theoretical $\nu_U$--$%
\nu_L$ relations for the black hole spacetime under consideration and
compare them with those of the Schwarzschild and RN black holes. The
resulting frequency correlations are presented and discussed in the
following section. 
\begin{figure*}[t!]
\centering
\includegraphics[width=0.28\linewidth]{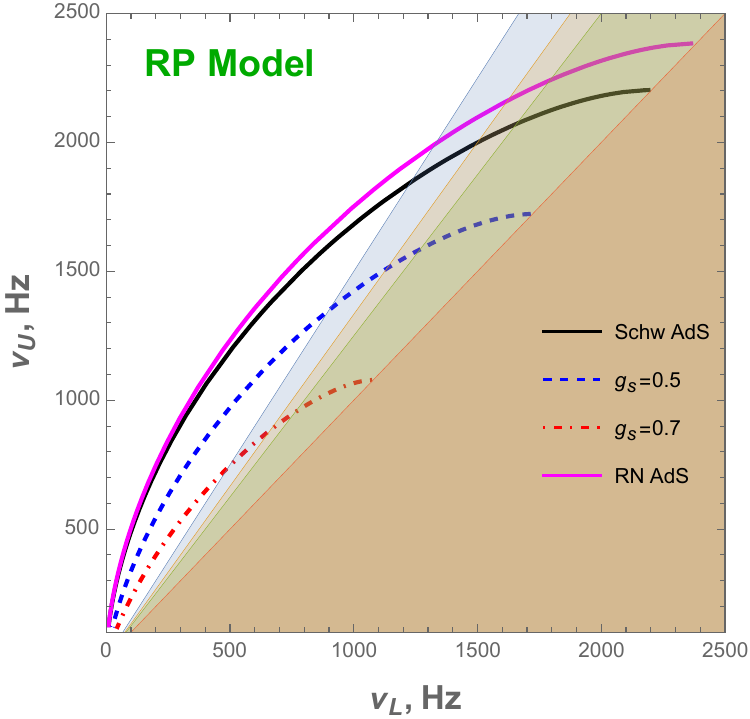} \includegraphics[width=0.28%
\linewidth]{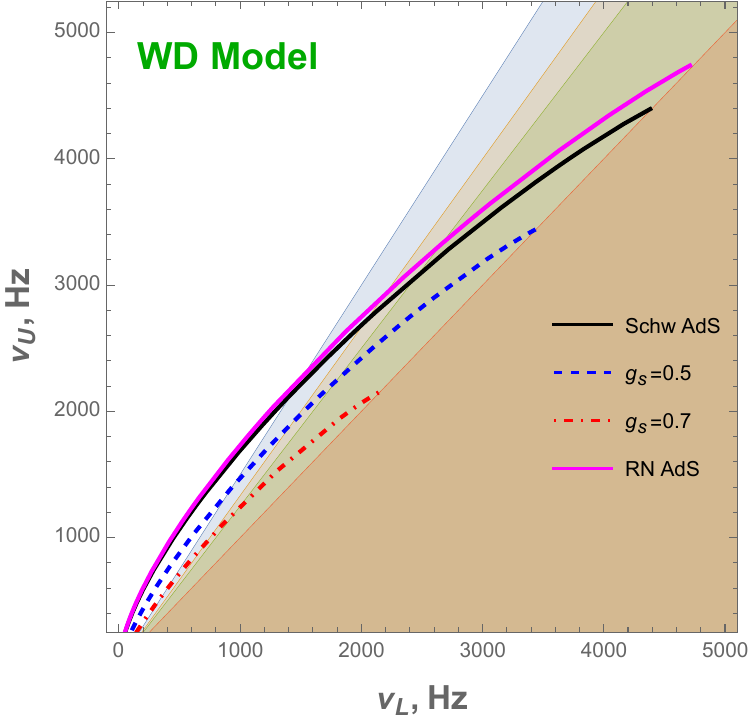} \includegraphics[width=0.28\linewidth]{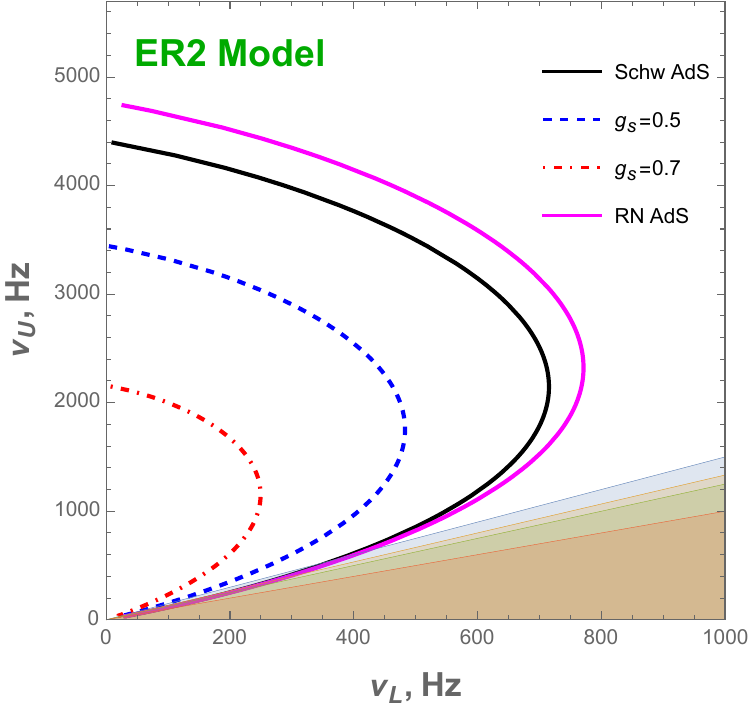} %
\includegraphics[width=0.28\linewidth]{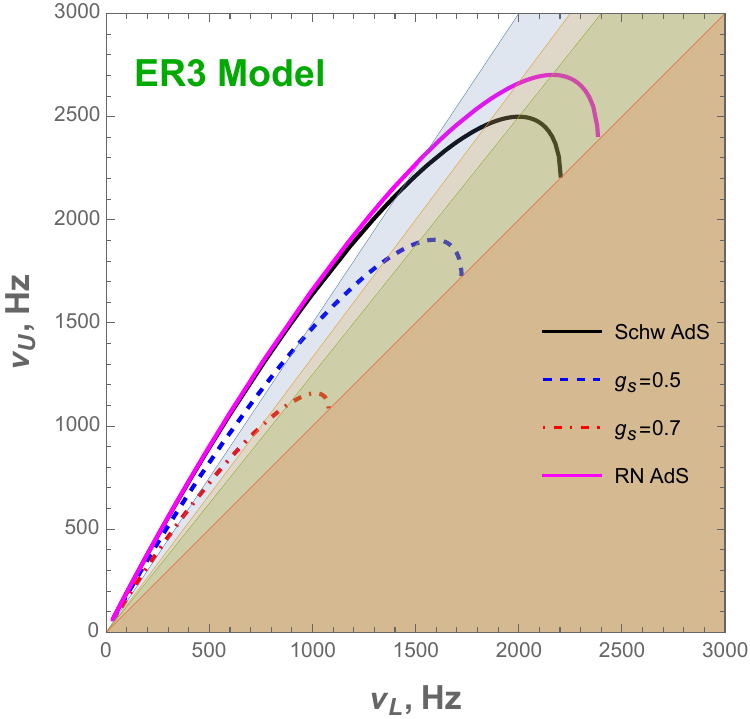} \includegraphics[width=0.28%
\linewidth]{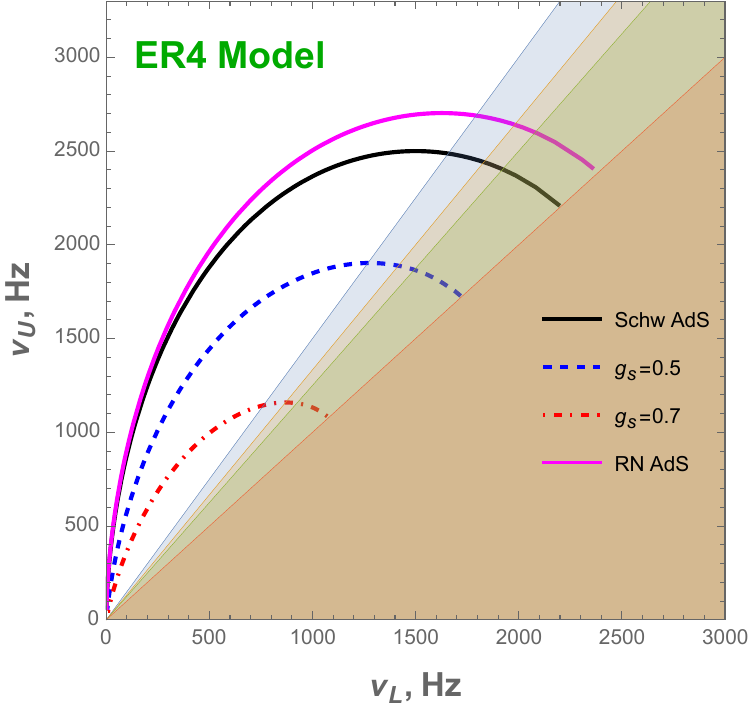}
\caption{ Radius of QPO orbits as a function of the coupling parameter $g_s$
in RP, WD, and ER2-4 models}
\label{figb2}
\end{figure*}

Figure~\ref{figb2} shows the correlations between the upper ($\nu_U$) and
lower ($\nu_L$) QPO frequencies predicted by each QPO models discussed. The
reference lines in the plots represents the frequency ratios of $3\!:\!2$, $%
4\!:\!3$, $5\!:\!4$, and $1\!:\!1$. The $1\!:\!1$ line is called the "QPO
graveyard." as it corresponds to cases where both QPO peaks merge.

\subsection{Markov Chain Monte Carlo Analysis}

To constrain the parameters of the proposed black hole solution, we perform
a Bayesian Markov Chain Monte Carlo (MCMC) analysis using QPO observations
from five well-studied black hole sources covering three distinct mass
scales. The sample consists of the stellar-mass black holes GRO J1655--40,
XTE J1550--564 and GRS 1915+105, the intermediate-mass black hole M87, and
the supermassive black hole Sgr A*. The observational properties of these
sources are summarized in Table~\ref{tab1b}. 
\begin{table*}[htb]
\centering
\begin{tabular}{|c|c|c|c|}
\hline
\textbf{Source} & \textbf{Mass} (in $M_\odot$) & \textbf{Upper Frequency (Hz)%
} & \textbf{Lower Frequency (Hz)} \\ \hline
GRO J1655$-$40 & $5.4 \pm 0.3$ \cite{t57} & $441 \pm 2$ \cite{t58} & $298
\pm 4$ \cite{t58} \\ \hline
XTE J1550$-$564 & $9.1 \pm 0.61$ \cite{t59} & $276 \pm 3$ & $184 \pm 5$ \\ 
\hline
GRS 1915$+$105 & $12.4^{+2.0}_{-1.8}$ \cite{t60} & $168 \pm 3$ & $113 \pm 5$
\\ \hline
M82 X-1 & $415 \pm 63$ \cite{nature} & $5.07 \pm 0.06$\cite{nature} & $3.32
\pm 0.06$\cite{nature} \\ \hline
Sgr A$^*$ & $(3.5 - 4.9) \times 10^6$ \cite{t64,t65} & $(1.445 \pm 0.16)
\times 10^{-3}$\cite{t66} & $(0.886 \pm 0.04) \times 10^{-3}$\cite{t66} \\ 
\hline
\end{tabular}%
\caption{Observational QPO data for different Black hole sources with
estimated mass \protect\cite{52}}
\label{tab1b}
\end{table*}

To illustrate the parameter estimation procedure, we employ the relativistic
precession (RP) model, which relates the observed twin-peak QPOs to the
orbital and radial epicyclic frequencies. Our choice of the RP model is
purely methodological and does not imply a preference over other QPO models.
Since the observed high-frequency QPOs frequently exhibit a frequency ratio
close to $3:2$, we also compare our results with those expected from
resonance-based models, where this ratio arises naturally.

The posterior probability distribution of the model parameters is obtained
using Bayes' theorem, 
\begin{equation}
P(\boldsymbol{\theta}|D,M)= \frac{P(D|\boldsymbol{\theta},M)\, \pi(%
\boldsymbol{\theta}|M)} {P(D|M)},  \label{eq:posterior}
\end{equation}
where $D$ denotes the observational data, $\pi(\boldsymbol{\theta}|M)$ is
the prior distribution, $P(D|\boldsymbol{\theta},M)$ is the likelihood
function, and $P(D|M)$ is the Bayesian evidence. The parameter vector is
chosen as 
\begin{equation}
\boldsymbol{\theta} = \left\{ M, q, g_s, l_s, \frac{r}{M} \right\}.
\end{equation}

Gaussian priors are assigned to all parameters, 
\begin{equation}
\pi(\theta_i) \propto \exp\left[ -\frac{1}{2} \left( \frac{%
\theta_i-\theta_{0,i}} {\sigma_i} \right)^2 \right], \qquad \theta_{\mathrm{%
low},i} < \theta_i < \theta_{\mathrm{high},i},
\end{equation}
where $\theta_{0,i}$ and $\sigma_i$ denote the mean and standard deviation
inferred from previous observational studies, while the parameter bounds are
imposed to ensure physically admissible solutions.

Assuming Gaussian measurement uncertainties, the total log-likelihood is
written as 
\begin{equation}
\log\mathcal{L} = \log\mathcal{L}_U + \log\mathcal{L}_L,
\end{equation}
where 
\begin{equation}
\log\mathcal{L}_U = -\frac12 \sum_i \frac{\left( \nu^{\mathrm{obs}}_{U,i} -
\nu^{\mathrm{th}}_{U,i} \right)^2} {\left( \sigma^{\mathrm{obs}}_{U,i}
\right)^2},
\end{equation}
and 
\begin{equation}
\log\mathcal{L}_L = -\frac12 \sum_i \frac{\left( \nu^{\mathrm{obs}}_{L,i} -
\nu^{\mathrm{th}}_{L,i} \right)^2} {\left( \sigma^{\mathrm{obs}}_{L,i}
\right)^2}.
\end{equation}
Here, $\nu_U^{\mathrm{obs}}$ and $\nu_L^{\mathrm{obs}}$ denote the observed
upper and lower QPO frequencies, respectively, while $\nu_U^{\mathrm{th}}$
and $\nu_L^{\mathrm{th}}$ are the corresponding theoretical frequencies
predicted by the RP model. \clearpage
\begin{figure}[]
\centering
\begin{subfigure}[b]{0.4\textwidth}
    \centering
    \includegraphics[width=\linewidth]{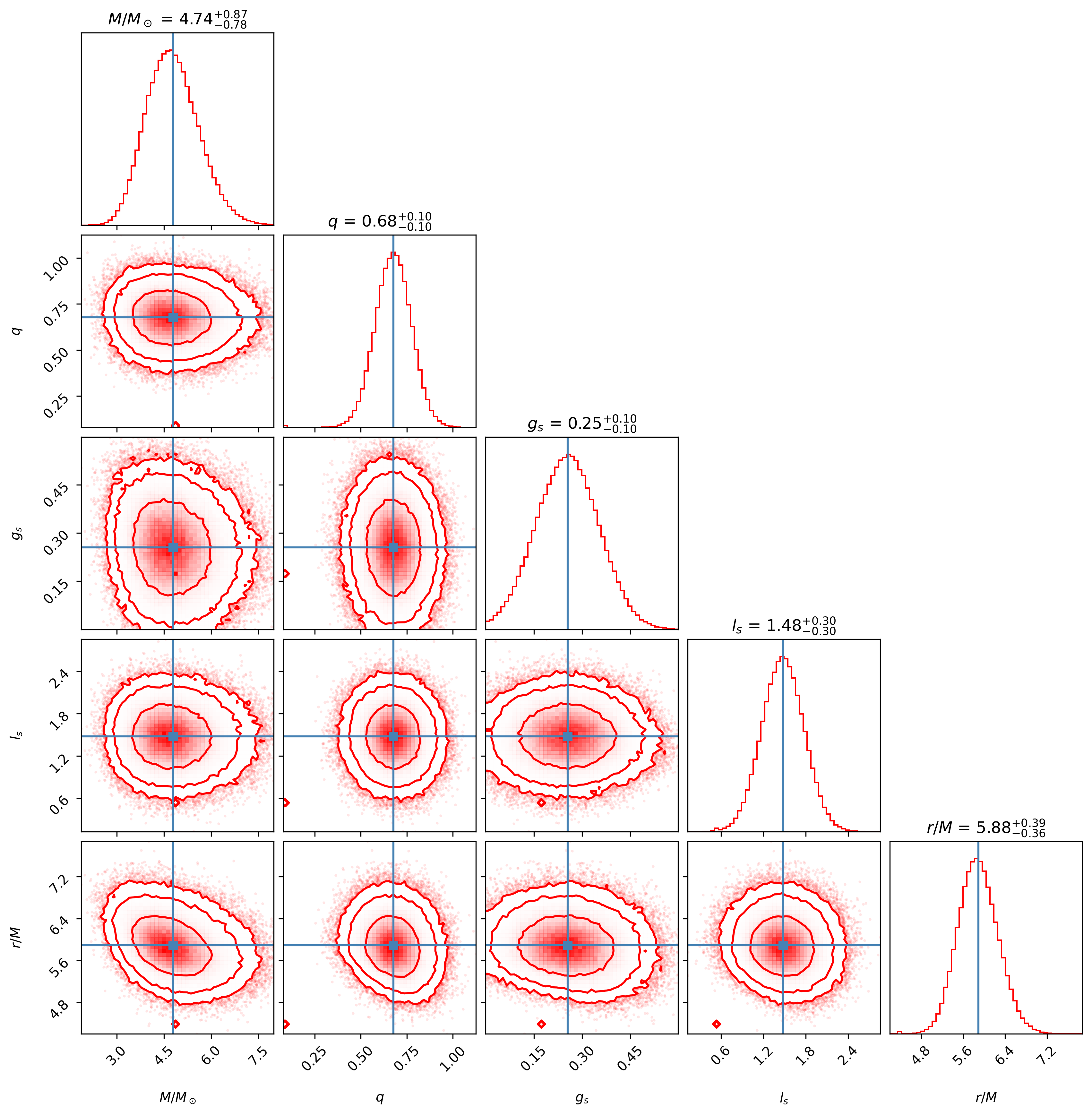}
    \caption{GRO J1655-40}
    \label{fig:rp}
\end{subfigure}
\hfill 
\begin{subfigure}[b]{0.4\textwidth}
    \centering
    \includegraphics[width=\linewidth]{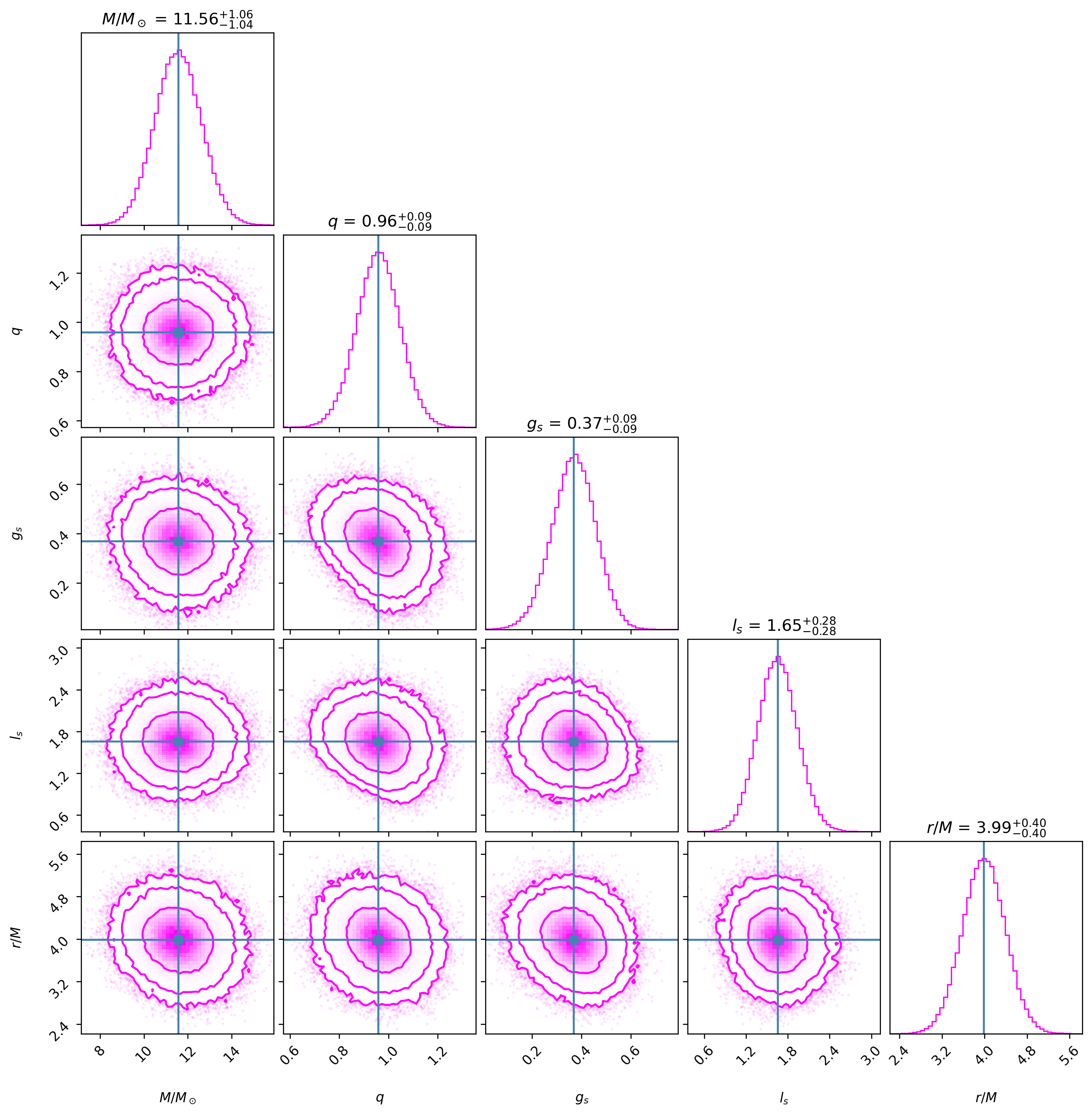}
    \caption{XTE J1550-564}
    \label{fig:wd}
\end{subfigure}
\hfill 
\begin{subfigure}[b]{0.4\textwidth}
    \centering
    \includegraphics[width=\linewidth]{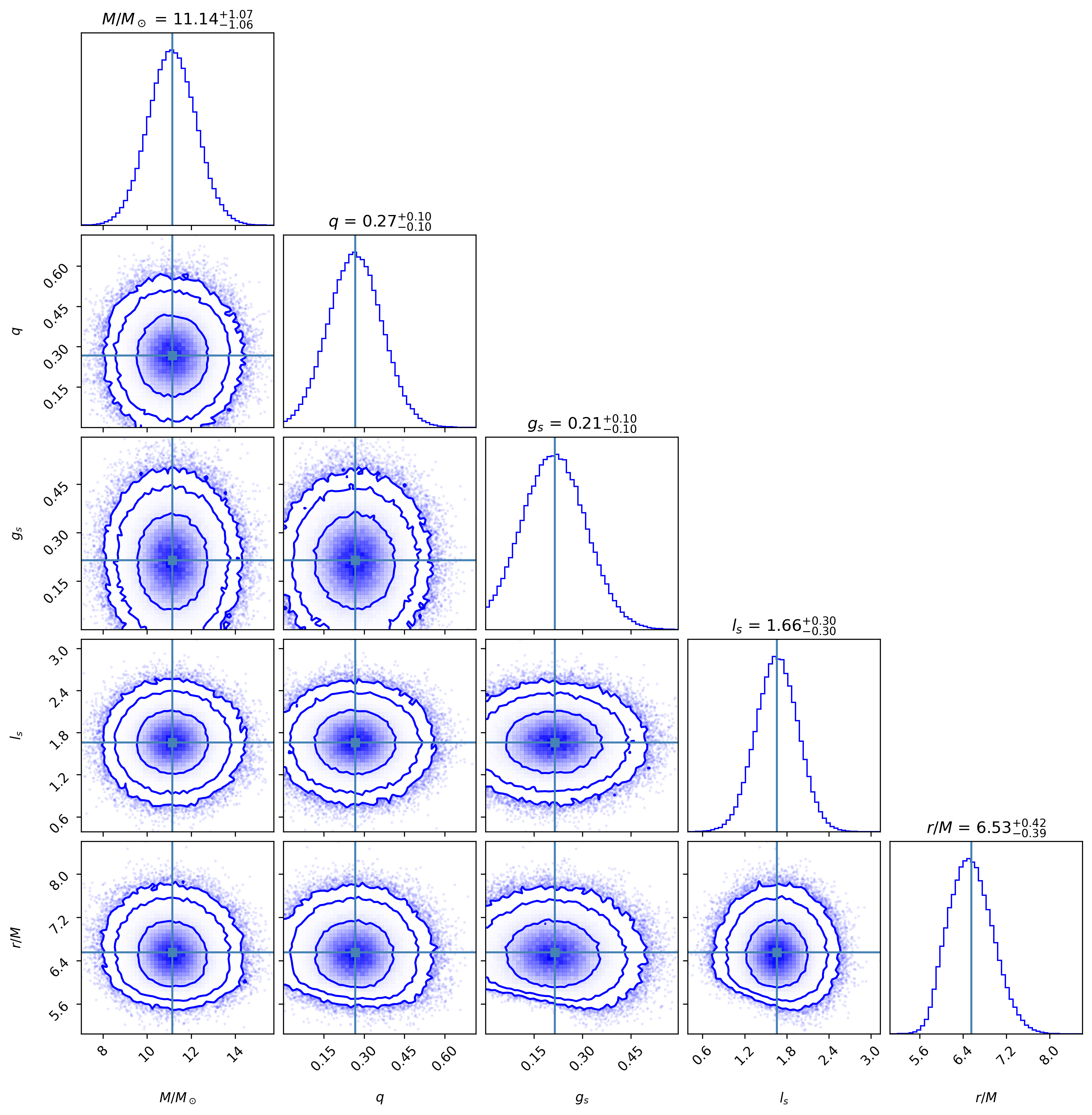}
    \caption{GRS 1915+105}
    \label{fig:er2}
\end{subfigure}
\begin{subfigure}[b]{0.4\textwidth}
    \centering
    \includegraphics[width=\linewidth]{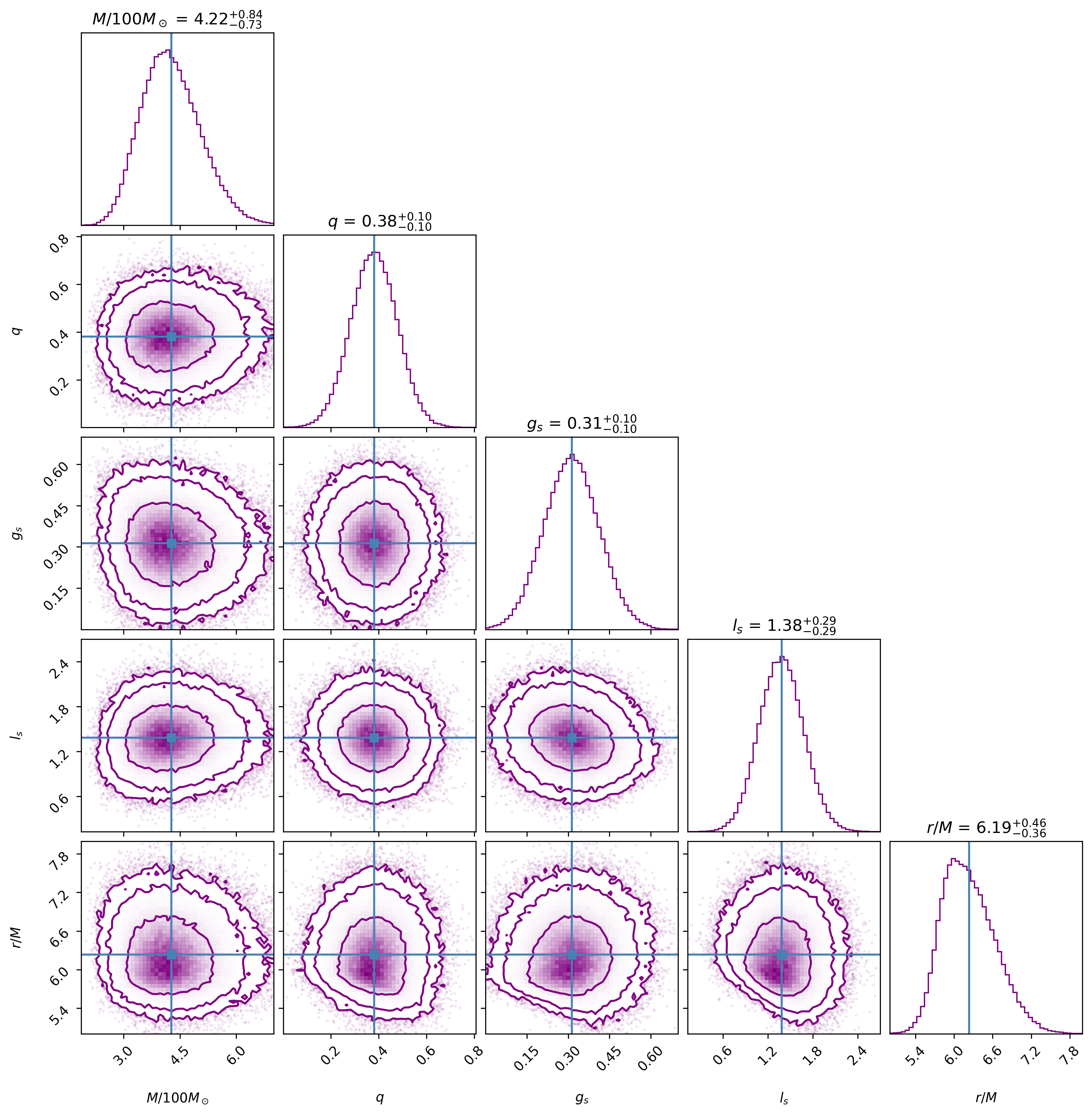}
    \caption{M82 X-1}
    \label{fig:er3}
\end{subfigure}
\hfill 
\begin{subfigure}[b]{0.4\textwidth}
    \centering
    \includegraphics[width=\linewidth]{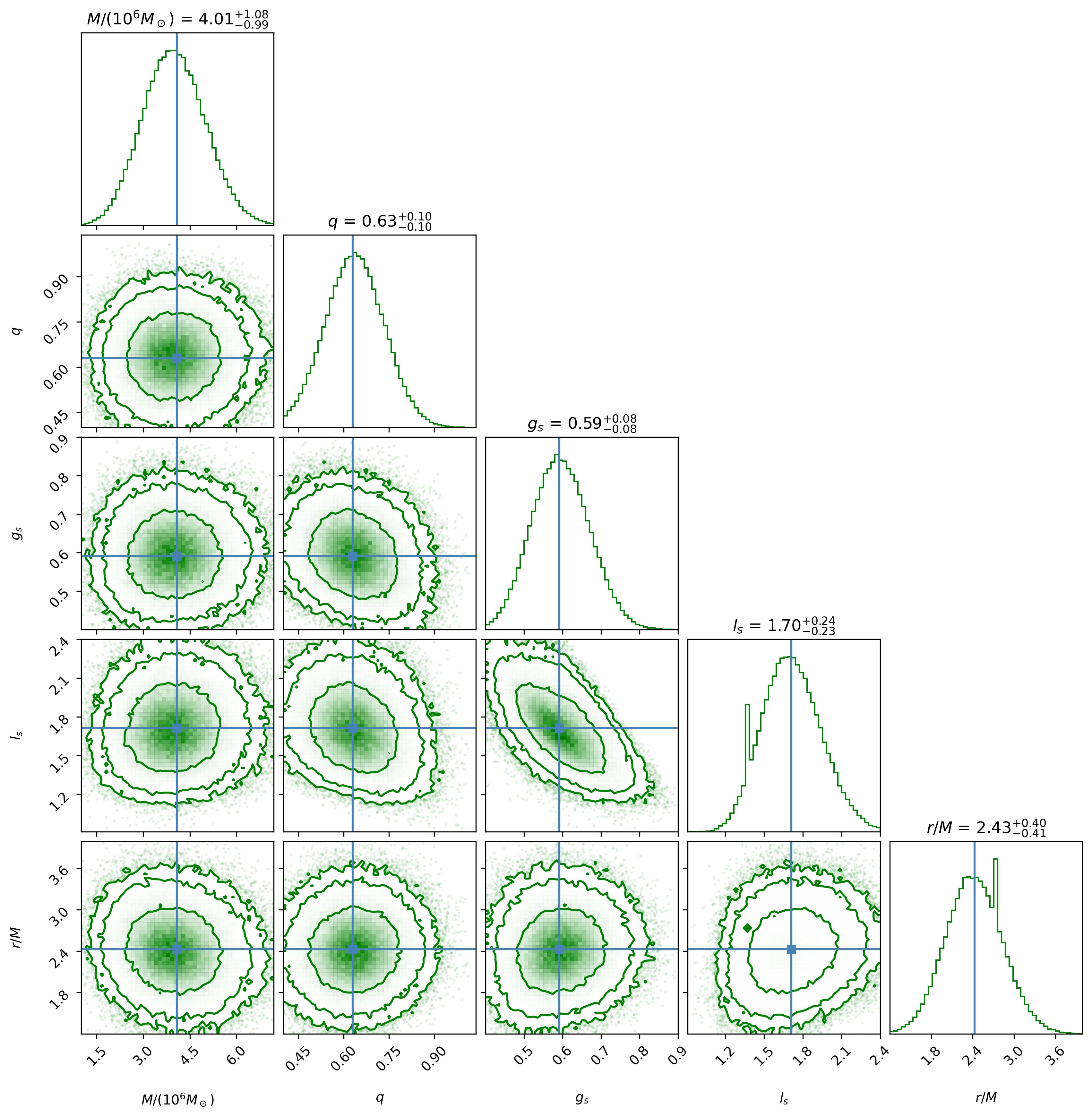}
    \caption{ Sgr A*}
    \label{fig:er4}
\end{subfigure}
\caption{Constraints on the parameters of the black hole solution using MCMC
analysis}
\label{figb3}
\end{figure}

The observational data employed in the analysis are listed in Table~\ref%
{tab1b}. Using these measurements together with the adopted priors, we
generate $10^5$ MCMC samples to explore the multidimensional parameter
space. The resulting posterior distributions allow us to determine the most
probable values and credible intervals of the black hole parameters $%
\{M,q,g_s,l_s,r/M\}$ that best reproduce the observed QPO frequencies.

Figure~\ref{figb3} presents the posterior distributions obtained from the
Markov Chain Monte Carlo (MCMC) analysis for the five model parameters,
namely the black hole mass $M$, the charge parameter $q$, the coupling
parameter $g_s$, the length of the string $l_s$, and the orbital radius $r/M$%
, for the five black hole sources considered in this work. The diagonal
panels show the marginalized one-dimensional posterior distributions, while
the off-diagonal panels display the corresponding two-dimensional joint
posterior distributions together with the $1\sigma$, $2\sigma$, and $3\sigma$
confidence contours.

For all five sources, the posterior distributions are well localized and
approximately Gaussian. The contours exhibit only weak to moderate
correlations among most of the parameters, suggesting that parameter
degeneracies are relatively small. The inferred black hole masses are
consistent with the corresponding observational estimates for stellar-mass,
intermediate-mass, and supermassive black holes. The normalized charge
parameter is constrained within the range $0.27 \lesssim q \lesssim 0.96$,
which means that the electric charge remains moderate for all sources.
Similarly, the coupling parameter is found to lie in the interval $0.21
\lesssim g_s \lesssim 0.59$, suggesting that only relatively small
corrections are required to reproduce the observed QPO frequencies. The
characteristic string length is consistently constrained to $1.38 \lesssim
l_s \lesssim 1.70$. The preferred orbital radius is determined to be $r/M
\simeq 2.43$--$6.53$. The observed twin-peak QPOs originate from regions
close to the innermost stable circular orbit, where relativistic effects
become significant.

The median values together with their corresponding $68\%$ credible
intervals are summarized in Table~\ref{tab2b}.

\begin{table*}[t]
\caption{Median values of the model parameters obtained from the MCMC
analysis together with their $68\%$ credible intervals.}
\label{tab2b}\centering
\begin{tabular}{lccccc}
\hline
Source & $M$ & $q$ & $g_s$ & $l_s$ & $r/M$ \\ \hline
GRO J1655--40 & $4.74^{+0.87}_{-0.78}\,M_\odot$ & $0.68^{+0.10}_{-0.10}$ & $%
0.25^{+0.10}_{-0.10}$ & $1.48^{+0.30}_{-0.30}$ & $5.88^{+0.39}_{-0.36}$ \\ 
XTE J1550--564 & $11.14^{+1.07}_{-1.06}\,M_\odot$ & $0.27^{+0.10}_{-0.10}$ & 
$0.21^{+0.10}_{-0.10}$ & $1.66^{+0.30}_{-0.30}$ & $6.53^{+0.42}_{-0.39}$ \\ 
M82 X--1 & $4.22^{+0.84}_{-0.73}\times10^{2}\,M_\odot$ & $%
0.38^{+0.10}_{-0.10}$ & $0.31^{+0.10}_{-0.10}$ & $1.38^{+0.29}_{-0.29}$ & $%
6.19^{+0.46}_{-0.36}$ \\ 
Sgr A* & $4.01^{+1.08}_{-0.99}\times10^{6}\,M_\odot$ & $0.63^{+0.10}_{-0.10}$
& $0.59^{+0.08}_{-0.08}$ & $1.70^{+0.24}_{-0.23}$ & $2.43^{+0.40}_{-0.41}$
\\ 
GRS 1915+105 & $11.56^{+1.06}_{-1.04}\,M_\odot$ & $0.96^{+0.09}_{-0.09}$ & $%
0.37^{+0.09}_{-0.09}$ & $1.65^{+0.28}_{-0.28}$ & $3.99^{+0.40}_{-0.40}$ \\ 
\hline
\end{tabular}%
\end{table*}

\newpage

\section{Conclusions}

In this study, exact black hole solutions are extracted for a static,
spherically symmetric spacetime incorporating a Letelier--Alencar \textit{%
cloud of strings} within Einstein--Maxwell--$\Lambda$ gravity.

The Ricci and Kretschmann scalars are found to diverge at $r=0$, thereby
indicating the presence of an essential curvature singularity at the origin.
Moreover, the spacetime is not asymptotically (A)dS, since the curvature
invariants do not approach their corresponding constant-\,$\Lambda$ limits
as $r\rightarrow\infty$, owing to the Letelier--Alencar \textit{cloud of
strings}.

The essential curvature singularity at $r=0$ is shown to be enclosed by the
event horizon, whose roots are systematically modified by the parameters $%
g_{s}$, $l_{s}$, and $q$. Below their respective critical thresholds, the
existence of both Cauchy and event horizons is verified, whereas naked
singularities or extremal horizons are produced in super-critical regimes.
Furthermore, the horizon radius is found to be maximized under
configurations of large $g_{s}$ paired with comparatively small values of $q$
and $l_{s}$.

Regarding the thermodynamic landscape, the conserved and thermodynamic
quantities are calculated, demonstrating their strict adherence to the first
law of thermodynamics. The Hawking temperature is shown to be dictated by
the complete parameter space, with its physical boundary ($T>0$) determined
by a unique root ($r_{+_{T=0}}$). Under sub-critical regimes of $g_s$, $l_s$%
, and $q$, two extrema are observed; these are subsequently suppressed in
super-critical regions, yielding a monotonic temperature growth.
Additionally, the physically viable thermodynamic domain of the black hole
is found to be systematically contracted upon increasing any of these three
parameters. Concurrently, a minimum mass threshold is shown to be governed
by $g_{s}$, $l_{s}$, and $q$, with its position shifted monotonically toward
larger values of both $M$ and $r_{+}$ as these parameters increase.

Concerning the thermodynamic stability, local and global behaviors are
rigorously assessed via the heat capacity and the Gibbs potential. By
analyzing the heat capacity, a small/large black hole phase transition is
shown to occur under subcritical values of $g_{s}$, $l_{s}$, and $q$. The
corresponding parameter space for local stability is found to be
systematically diminished as these parameters increase. Similarly, the
global thermodynamic stability, characterized by the negative regions of the
Gibbs potential $G$, is restricted by larger values of $g_{s}$, $l_{s}$, and 
$q$. Under subcritical regimes, $G$ exhibits multi-root structures that
demarcate distinct stable and unstable phases, whereas supercritical regimes
restrict global stability primarily to large black holes. Crucially, large
black hole configurations are found to consistently satisfy both local and
global stability criteria, demonstrating an excellent thermodynamic
alignment.

To connect these theoretical results with observations, the parameters of
the proposed black hole model are constrained using the observed twin-peak
QPO frequencies through a Bayesian MCMC analysis. Three different classes of
black holes, namely stellar-mass, intermediate-mass, and supermassive black
holes, are considered. The posterior distributions for all five analyzed
sources are found to be well localized and approximately Gaussian. The
inferred black hole masses are observed to be in good agreement with the
corresponding observational estimates across all three mass scales.
Specifically, the normalized charge parameter is constrained to the range $%
0.27 \lesssim q \lesssim 0.96$, the coupling parameter is restricted to $%
0.21 \lesssim g_s \lesssim 0.59$ (indicating that only modest corrections
are required to reproduce the observed QPO frequencies), and the
characteristic string length is consistently found within the interval $1.38
\lesssim l_s \lesssim 1.70$. Moreover, the preferred orbital radius is
located in the range $2.43 \lesssim r/M \lesssim 6.53$. These observational
constraints indicate that the observed twin-peak QPOs originate from regions
in close proximity to the innermost stable circular orbit (ISCO), where
strong-field relativistic effects dominate the particle dynamics. Overall,
consistent constraints on the black hole parameters are successfully
provided by the MCMC analysis.

Finally, it should be emphasized that the observed QPO data used in this
analysis originate from astrophysical black holes, which are generally
expected to be rotating. In contrast, the black hole solution considered in
this work is static and spherically symmetric. Consequently, the parameters
constrained through the MCMC analysis should be regarded as effective
parameters that reproduce the observed QPO frequencies within the framework
of the static model. Since the observational frequencies intrinsically
contain the effects of black hole spin, the inferred parameter values may
effectively incorporate contributions arising from rotation. Therefore, the
present constraints should not be interpreted as purely intrinsic parameters
of the static solution. For a more rigorous determination of the physical
parameters, the rotating counterpart of the present black hole is required.
This counterpart can be constructed using the Newman--Janis algorithm, which
is significantly more challenging from both analytical and computational
perspectives. Owing to these complexities, a comprehensive investigation of
rotating configurations is left for future work.

\acknowledgements{B. Eslam Panah thanks the University of Mazandaran.}


\begin{thebibliography}{999}
\bibitem{LIGO2016} B. P. Abbott et al. (LIGO Scientific, Virgo), Phys. Rev.
Lett. \textbf{116}, 061102 (2016).

\bibitem{LIGO2017} B. P. Abbott et al. (LIGO Scientific, Virgo), Phys. Rev.
Lett. \textbf{119}, 141101 (2017).

\bibitem{LIGO2020} R. Abbott et al. (LIGO Scientific, Virgo), Astrophys. J.
Lett. \textbf{896}, L44 (2020).

\bibitem{ETH2019} K. Akiyama et al. (Event Horizon Telescope), Astrophys. J.
Lett. \textbf{875}, L1 (2019).

\bibitem{ETH2022} K. Akiyama et al. (Event Horizon Telescope), Astrophys. J.
Lett. \textbf{930}, L12 (2022).

\bibitem{Medved2004} A. J. M. Medved, D. Martin, and M. Visser, Class.
Quant. Grav. \textbf{21}, 1393 (2004).

\bibitem{Narayan2005} R. Narayan, New J. Phys. \textbf{7}, 199 (2005).

\bibitem{Cunha2015} P. V. P. Cunha, C. A. R. Herdeiro, E. Radu, and H. F.
Runarsson, Phys. Rev. Lett. \textbf{115}, 211102 (2015).

\bibitem{Konoplya2021} R. A. Konoplya, Phys. Lett. B \textbf{823}, 136734
(2021).

\bibitem{Cardoso2022} V. Cardoso, et al., Phys. Rev. D \textbf{105}, L061501
(2022).

\bibitem{Rahmatov2025} B. Rahmatov, et al., Phys. Dark Univ. \textbf{50},
102152 (2025).

\bibitem{Hoshimov2025} H. Hoshimov, S. Orzuev, F. Atamurotov, and A.
Abdujabbarov, Annals Phys. \textbf{482}, 170209 (2025).

\bibitem{Bakhodirov2025} A. Bakhodirov, et al., Phys. Dark Univ. \textbf{49}%
, 102008 (2025).

\bibitem{Barausse2014} E. Barausse, V. Cardoso, and P. Pani, Phys. Rev. D 
\textbf{89}, 104059 (2014).

\bibitem{Macedo2016} C. F. B. Macedo, L. C. S. Leite, and L. C. B. Crispino,
Phys. Rev. D \textbf{93}, 024027 (2016).

\bibitem{Macedo2024} C. F. B. Macedo, J. L. Rosa, and D. Rubiera-Garcia,
JCAP \textbf{07}, 046 (2024).

\bibitem{Letelier1979} P. S. Letelier, Phys. Rev. D \textbf{20}, 1294 (1979).

\bibitem{Rincon2018} A. Rincon, and G. Panotopoulos, Eur. Phys. J. C \textbf{%
78}, 858 (2018).

\bibitem{Li2021} Z. Li, and T. Zhou, Phys. Rev. D \textbf{104}, 104044
(2021).

\bibitem{Belhaj2022} A. Belhaj, and Y. Sekhmani, Gen. Rel. Grav. \textbf{54}%
, 17 (2022).

\bibitem{He2022} A. He, J. Tao, Y. Xue, and L. Zhang, Chin. Phys. C \textbf{%
46}, 065102 (2022).

\bibitem{Toledo2019} J. M. Toledo, and V. B. Bezerra, Eur. Phys. J. C 
\textbf{79}, 110 (2019).

\bibitem{Singh2020} D. V. Singh, S. G. Ghosh, and S. D. Maharaj, Phys. Dark
Univ. \textbf{30}, 100730 (2020).

\bibitem{Rodrigues2022b} M. E. Rodrigues, and M. V. d. S. Silva, Phys. Rev.
D \textbf{106}, 084016 (2022).

\bibitem{Sadeghi2024} J. Sadeghi, S. N. Gashti, I. Sakalli, and B.
Pourhassan, Nucl. Phys. B \textbf{1004}, 116581 (2024).

\bibitem{Ghosh2014} S. G. Ghosh, U. Papnoi, and S. D. Maharaj, Phys. Rev. D 
\textbf{90}, 044068 (2014).

\bibitem{Toledo2018} J. de M. Toledo, and V. B. Bezerra, Eur. Phys. J. C 
\textbf{78}, 534 (2018).

\bibitem{Waseem2023} A. Waseem, F. Javed, M. Z. Gul, G. Mustafa, and A.
Errehymy, Eur. Phys. J. C \textbf{83}, 1088 (2023).

\bibitem{Rodrigues2022} M. E. Rodrigues, and H. A. Vieira, Phys. Rev. D 
\textbf{106}, 084015 (2022).

\bibitem{Muniz2025} C. R. Muniz, et al., Phys. Dark Univ. \textbf{52},
102272 (2026).

\bibitem{Mustafa2022} G. Mustafa, et al., Chin. Phys. C \textbf{46}, 125107
(2022).

\bibitem{Atamurotov2022} F. Atamurotov, I. Hussain, G. Mustafa, and K.
Jusufi, Eur. Phys. J. C \textbf{82}, 831 (2022).

\bibitem{Alencar} G. Alencar, R. R. Landim, and R. N. Costa Filho, Phys.
Dark Univ. \textbf{49}, 102031 (2025).

\bibitem{Silva2026} M. V. de S. Silva, et al., Phys. Rev. D \textbf{113},
064052 (2026).

\bibitem{17} L. Angelini, L. Stella, and A. N. Parmar, Astrophys. J. \textbf{%
346}, 906 (1989).

\bibitem{18} S. Kato, and J. Fukue, Publ. Astron. Soc. Jpn. \textbf{32}, 377
(1980).

\bibitem{19} M. A. Abramowicz, and W. Kluzniak, Astron. Astrophys. \textbf{%
374}, L19 (2001).

\bibitem{20} R. V. Wagoner, A. S. Silbergleit, and M. Ortega-Rodriguez,
Astrophys. J. \textbf{559}, L25 (2001).

\bibitem{21} A. S. Silbergleit, R. V. Wagoner, and M. Ortega-Rodriguez,
Astrophys. J. \textbf{548}, 335 (2001).

\bibitem{22} D. H. Wang, et al., Mon. Not. R. Astron. Soc. \textbf{454},
1231 (2015).

\bibitem{23} L. Rezzolla, S. I. Yoshida, T. J. Maccarone, and O. Zanotti,
Mon. Not. R. Astron. Soc. \textbf{344}, L37 (2003).

\bibitem{24} G. Torok, and Z. Stuchlik, Astron. Astrophys. \textbf{437}, 775
(2005).

\bibitem{25} A. Ingram, and C. Done, Mon. Not. R. Astron. Soc. \textbf{405},
2447 (2010).

\bibitem{26} P. C. Fragile, O. Straub, and O. Blaes, Mon. Not. R. Astron.
Soc. \textbf{461}, 1356 (2016).

\bibitem{27} Z. Stuchlik, A. Kotrlova, and G. Torok, Astron. Astrophys. 
\textbf{552}, A10 (2013).

\bibitem{28} Z. Stuchlik, A. Kotrlova, and G. Torok, Acta Astron. \textbf{62}%
, 389 (2012).

\bibitem{29} M. Ortega-Rodriguez, H. Solis-Sanchez, L. Alvarez-Garcia, and
E. Dodero-Rojas, Mon. Not. R. Astron. Soc. \textbf{492}, 1755 (2020).

\bibitem{30} Z. Stuchlik, M. Kolos, J. Kovar, P. Slany, and A. Tursunov,
Universe. \textbf{6}, 26 (2020).

\bibitem{31} A. Maselli, et al., Astrophys. J. \textbf{899}, 139 (2020).

\bibitem{32} J. Rayimbaev, K.F. Dialektopoulos, F. Sarikulov, and A.
Abdujabbarov, Eur. Phys. J. C \textbf{83}, 572 (2023).

\bibitem{33} O. Donmez, Appl. Math. Comput. \textbf{181}, 256 (2006).

\bibitem{34} O. Donmez, Res. Astron. Astrophys. \textbf{24}, 085001 (2024).

\bibitem{35} O. Donmez, Eur. Phys. J. C \textbf{84}, 524 (2024).

\bibitem{36} O. Donmez, Mod. Phys. Lett. A \textbf{39}, 2450076 (2024).

\bibitem{37} F. Koyuncu, and O. Donmez, Mod. Phys. Lett. A \textbf{29},
1450115 (2014).

\bibitem{39} O. Donmez, Phys. Lett. B \textbf{827}, 136997 (2022).

\bibitem{42} O. Donmez, J. High Energy Astrophys. \textbf{45}, 1 (2025).

\bibitem{43} O. Donmez, and F. Dogan, Phys. Dark Univ. \textbf{46}, 101718
(2024).

\bibitem{c1} A. Dasgupta, N. Tiwari and I. Banerjee, JCAP \textbf{10} 054
(2025).

\bibitem{c2} I. Banerjee, JCAP \textbf{08}, 034 (2022).

\bibitem{c3} I. Banerjee, S. Chakraborty and S. SenGupta, JCAP \textbf{09},
037 (2021).

\bibitem{c4} S. Jumaniyozov, et al., Eur. Phys. J. C \textbf{85}, 126 (2025).

\bibitem{c5} J. Rayimbaev, S. Murodov, A. Shermatov and A. Yusupov, Eur.
Phys. J. C \textbf{84}, 1114 (2024).

\bibitem{c6} E. Ghorani, et al., Eur. Phys. J. C \textbf{84}, 1022 (2024).

\bibitem{c7} S. Jumaniyozov, et al., Eur. Phys. J. C \textbf{84}, 964 (2024).

\bibitem{local1} R. G. Cai, and A. Wang, Phys. Rev. D \textbf{70}, 064013
(2004).

\bibitem{local2} H. A. Gonzalez, M. Hassaine, and C. Martinez, Phys. Rev. D 
\textbf{80}, 104008 (2009).

\bibitem{local3} O. Gurtug, S. H. Mazharimousavi, M. Halilsoy, Phys. Rev. D 
\textbf{85}, 104004 (2012).

\bibitem{local4} M. S. Ma, and R. Zhao, Phys. Lett. B \textbf{751}, 278
(2015).

\bibitem{local5} S. Fernando, Phys. Rev. D \textbf{94}, 124049 (2016).

\bibitem{local6} B. Eslam Panah, M. E. Rodrigues, Eur. Phys. J. C \textbf{83}%
, 237 (2023).

\bibitem{local7} B. Eslam Panah, Phys. Lett. B \textbf{844}, 138111 (2023).

\bibitem{local8} B. Eslam Panah, M. E. Rodrigues, Eur. Phys. J. C \textbf{84}%
, 1125 (2024).

\bibitem{local9} B. Eslam Panah, Phys. Lett. B \textbf{868}, 139711 (2025).

\bibitem{Hawking:1983} S. Hawking, and D. N. Page, Commun. Math. Phys. 
\textbf{87}, 577 (1983).

\bibitem{Gibbs1} S. Fernando, Phys. Rev. D \textbf{74}, 104032 (2006).

\bibitem{Gibbs2} M. Dehghani, and M. R. Setare, Phys. Rev. D \textbf{100},
044022 (2019).

\bibitem{Gibbs3} L. Balart, and S. Fernando, Mod. Phys. Lett. A \textbf{41},
2650059 (2026).

\bibitem{Gibbs4} B. Eslam Panah, B. Hamil, and M. E. Rodrigues, Prog. Theor.
Exp. Phys. \textbf{2026}, 053E03 (2026).

\bibitem{Gibbs5} B. Eslam Panah, B. Hamil, M. E. Rodrigues, Eur. Phys. J. C 
\textbf{86}, 81 (2026).

\bibitem{Bambi2017book} C. Bambi, \textit{Black Holes: A Laboratory for
Testing Strong Gravity} (2017).

\bibitem{51} M. Shahzadi, M. Kolos, Z. Stuchlik, Y. Habib, Eur. Phys. J. C 
\textbf{81}, 1067 (2021).

\bibitem{StellaVietri1998} L. Stella, and M. Vietri, Astrophys. J. Lett. 
\textbf{492},~L59 (1998).

\bibitem{StellaVietri1999} L. Stella, and M. Vietri, Phys. Rev. Lett. ~%
\textbf{82}, 17 (1999).

\bibitem{MorsinkStella1999} S. M. Morsink, and L.~Stella, Astrophys. J. 
\textbf{513}, 827 (1999).

\bibitem{wd1} S. Kato, Publ. Astron. Soc. Jap. \textbf{56}, 559 (2004).

\bibitem{wd2} S. Kato, Publ. Astron. Soc. Jap. \textbf{56}, 905 (2004).

\bibitem{52} I. Banerjee, JCAP \textbf{05}, 020 (2022).

\bibitem{t57} M. E. Beer, and P. Podsiadlowski, Mon. Not. Roy. Astron. Soc. 
\textbf{331}, 351 (2002).

\bibitem{t58} S. E. Motta, et al., Mon. Not. Roy. Astron. Soc. \textbf{437},
3 (2014).

\bibitem{t59} J. A. Orosz, et al., Astrophys. J. \textbf{730}, 75 (2011).

\bibitem{t60} M. J. Reid, et al., Astrophys. J. \textbf{796}, 2 (2014).

\bibitem{nature} D. R Pasham, T. E Strohmayer, and R. F. Mushotzky, Nature. 
\textbf{513}, 74 (2014).

\bibitem{t64} A. M. Ghez et al., Astrophys. J. \textbf{689}, 1044 (2008).

\bibitem{t65} S. Gillessen, et al., Astrophys. J. \textbf{692}, 1075 (2009).

\bibitem{t66} Z. Stuchlik and A. Kotrlova, Gen. Rel. Grav. \textbf{41}, 1305
(2009).
\end{thebibliography}
\end{document}